%% file: Varna_paper.tex
\documentclass[12pt]{article}
\usepackage{epsf}
\input CrayolaColors.tex

\newcommand{\dNuphi}{\partial^{\nu}\phi} 
\newcommand{\dLdmuphi}{\partial{\cal L}\over\partial(\partial_{\mu}\phi)}

\newcommand{\cd}{\makebox[0.08cm]{$\cdot$}}
\newcommand{\be}{\begin{equation}} \newcommand{\ee}{\end{equation}}

\begin{document}

{\Large \bf Two-Body Bound States in Light-Front Dynamics}\bigskip

\vspace{0.5cm}
{\bf J.Carbonell$^{(a)}$, M. Mangin-Brinet$^{(a)}$\footnote{Now at Geneva University and CERN, Switzerland}, V.A. Karmanov$^{(b)}$}
\bigskip

\begin{small}
{$^{(a)}$ Institut des  Sciences Nucl\'{e}aires, 53 avenue des Martyrs, 38026 Grenoble Cedex, France}

{$^{(b)}$ Lebedev Physical Institute, Leninsky Prospect 53, 119991 Moscow, Russia}
\end{small}

\vspace{0.5cm}
\begin{abstract}
We present the main features of the explicitly covariant Light-Front Dynamics
formalism  and a summary of our recent works on this topic.
They concern the bound states of two scalar particles in the Wick-Cutkosky
model and of two fermions interacting via the usual OBEP ladder kernels.
\end{abstract}
\hspace{.5cm}

\section{Motivation}

The recent measurements performed at CEBAF/TJNAF on the deuteron
structure functions and tensor polarization up to
momentum transfer values yet never reached, was the starting
motivation for a series of works in relativistic dynamics. Indeed the measurement  of the electromagnetic form
factors at values $Q^2\approx$ 6 (GeV/c)$^2$, much greater than
the nucleon mass, lets no hope for any attempt to describe this
physics without relativity. There are even very few chances that
meson-nucleon dynamics could still be used while the distances
involved are smaller than the nucleon itself $R_N\approx$0.86 fm.
A recent and complete reviews of both the experimental and theoretical
 works can be found in \cite{GVO_Revue_01,Gross} and references therein.

This motivation has always been present in theoretical physics since the first days of quantum
mechanics.
It relies on the need for having a proper relativistic wavefunction of  simple systems -- an
essential ingredient in the N\={N}, q\={q}, qqq, \ldots physics -- and for clarifying the
kind of  modifications that Lorentz invariance brings into the non relativistic dynamics.

Our first contribution in this field consisted in a series of
papers \cite{CK_NPA581_95,CK_NPA589_95,CDMK_98,CK_EPJA_99} which,
in the framework of the Explicitly Covariant Light-Front Dynamics
(ECLFD), led us to predict the deuteron structure functions and
polarization observables. These calculations, though performed in
a perturbative way over the wavefunctions of the Bonn model
\cite{Bonn}, turned to be quite successful in the deuteron case
and have recently found a natural application in describing the NN
correlation functions in nuclei (see Prof. Antonov
contribution to this School and \cite{Antonov_02}). The success
encountered in these first works and some clear advantages of the
ECLFD itself in describing relativistic composite systems,
convinced us to go deeper inside this approach both in developing
formal aspects of the theory and in studying in detail the
properties of its  solutions. This effort has up to now resulted
in studying the bound states solutions of two scalar and two
fermions systems in the ladder approximation. Doing so, we have in
mind the description of  "genuine relativistic systems" (those for
which $B\sim m$, $\beta\sim 1$) as well as  the {\it a priori}
 non relativistic ones when probed in relativistic regions (e.g. deuteron).

In this contribution, we first present
the leading ideas of the ECLFD formalism (Section \ref{Formalism}) and
then summarize the main results of our recent works \cite{MC_PLB_00}-\cite{MMB_These_01}
on this topic, both in the scalar (Section \ref{Scalars})
and fermionic (Section \ref{Fermions}) case.
We put special emphasis in the following items:
i) the size and nature of relativistic effects
ii) the comparison with the non relativistic solutions and other relativistic approaches
iii) the problem of constructing the non zero angular momentum states and
iv) the solution of the OBEP models for fermions and their stability properties
with respect the cutoff.

\section{The formalism}\label{Formalism}

Light-Front Dynamics is an hamiltonian formulation of the quantum
field  theory, specially well adapted to the description of
relativistic composite systems. In its explicitly covariant
version, proposed by V.A. Karmanov \cite{VAK}  and recently reviewed in \cite{CDMK_98,VAK_Mitra},
the state vector is defined on a space-time hyperplane whose
equation is given by $\omega\cdot x=\sigma$ with $\omega^2=0$.
Wavefunctions - defined as the Fock components of the state vector
- are the usual formal objects of this theory and are directly
comparable to their non relativistic counterparts.

As in any hamiltonian formalism, the dynamical equations
are provided by the eigenvalue equations of the two Casimir operators of
the Poincar\'e group,
respectively  the square of the four-momentum $\hat{P}^{\mu} $
and of the Pauli-Lubanski $\hat{S}_{\mu}$ four-vectors
\begin{eqnarray}
\hat{P}^2 \mid\Psi> &=&  M^2 \mid\Psi> \label{P2} \\
\hat{S}^2 \mid\Psi> &=& -M^2 J(J+1) \mid\Psi> \label{S2}
\end{eqnarray}
These operators are quadratic forms of the ten generators
$\left\{\hat{P}_{\mu},\hat{J}_{\rho \sigma}\right\}$ of the Poincar\'e group:
\begin{eqnarray*}
\hat{P}^2    &=& \hat{P}_{\mu}\hat{P}^{\mu}  \\
\hat{S}_{\mu}&=& {1\over2}\epsilon_{\mu \nu \rho \sigma} \hat{P}^{\nu} \hat{J}^{\rho\sigma}
\end{eqnarray*}

\begin{figure}[hbtp]
\begin{center}
\epsfxsize=7.cm \mbox{\epsffile{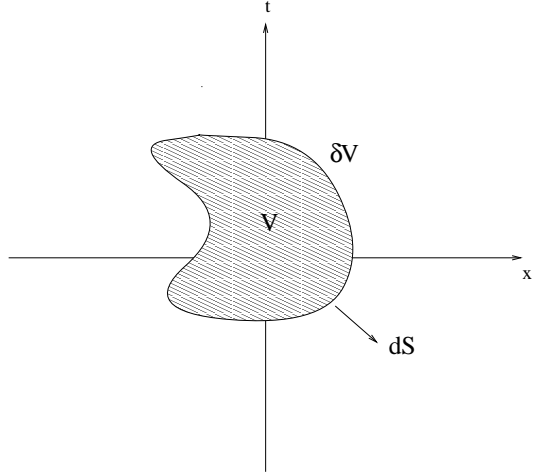}}
\caption{Arbitrary volume $V$ for integrating continuity equation}\label{FluxVqconque}
\end{center}
\end{figure}

Different ways of constructing the generators set would lead to different relativistic theories
\cite{Dirac_RMP_49}
which will be furthermore strictly equivalent if they were solved in their full complexity.
This choice is provided by integrating the Noether conserved currents over a suitable 3D space-time surface.
We remind that Noether theorem associates to each continuous space-time transformation
\begin{equation}\label{Lorentz}
 x\rightarrow x'=\Lambda x + a
\end{equation}
letting invariant a lagrangian density
\begin{equation}\label{Ldensity}
{\cal L}(x)={\cal L}_0(x) + {\cal L}_{int}(x)
\end{equation}
a  4-current $j^{\mu}$ satisfying the continuity equation
\begin{equation}\label{dmujmu}
\partial_{\mu}j^{\mu}=0
\end{equation}
This latter is calculated from a particular ${\cal L}(x)$
according to very precise recipes. Integrating over an arbitrary
4D volume $V$ (figure \ref{FluxVqconque}), equation
$(\ref{dmujmu})$ is equivalent to assert the nullity of flux across its boundary $\partial V$
\begin{equation} \label{FN}
 \int_{\partial V} j^{\mu} dS_{\mu}=0
\end{equation}
where $dS^{\mu}$ denotes the normal to $\partial V$.

\begin{figure}[hbtp]
\begin{center}
\begin{minipage}[t]{80mm}
\epsfxsize=6.cm\mbox{\epsffile{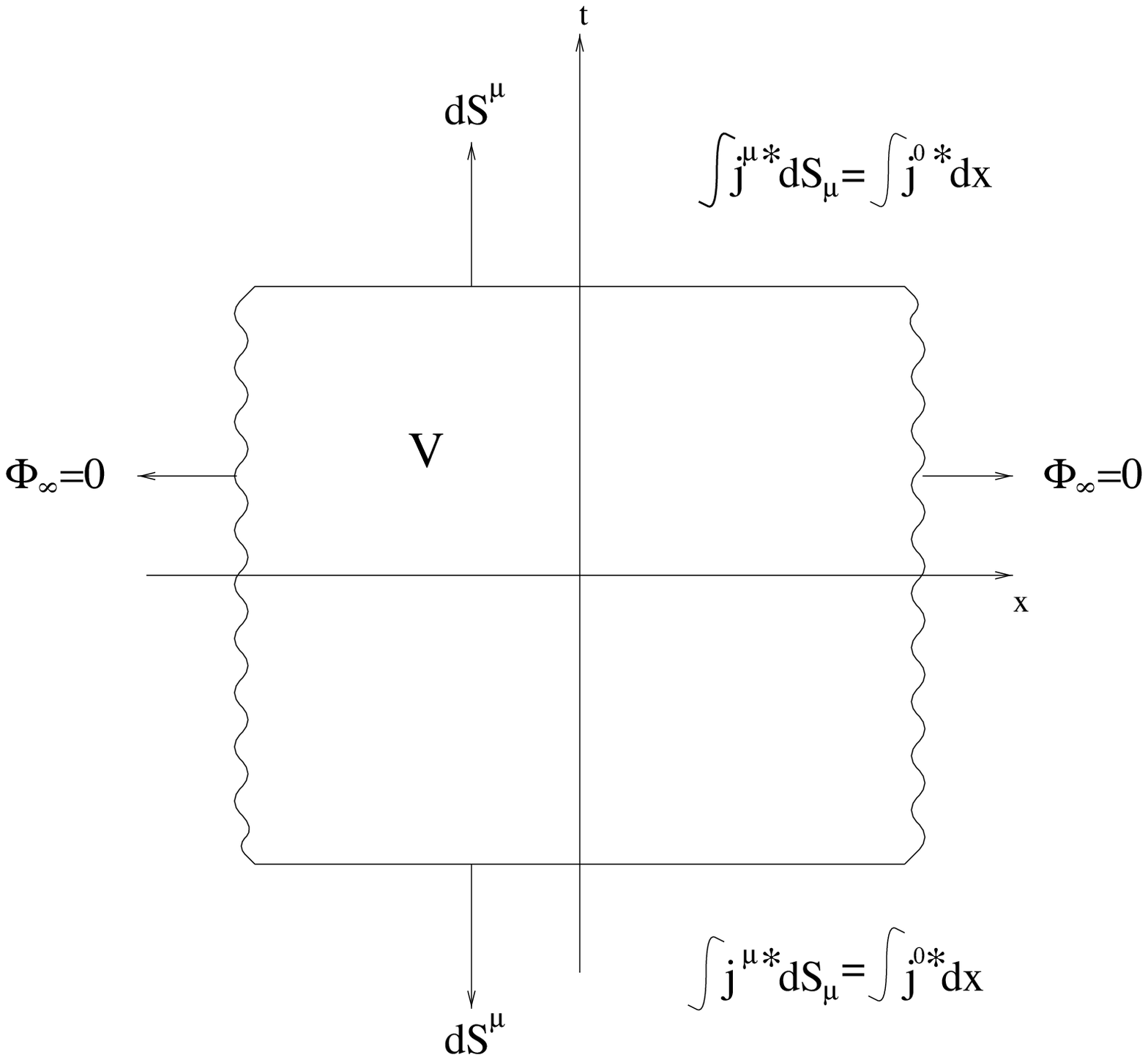}}
\caption{Flux in instant-form of dynamics}\label{flux_xt}
\end{minipage}
\hspace{0.3cm}
\begin{minipage}[t]{80mm}
\epsfxsize=6.cm\mbox{\epsffile{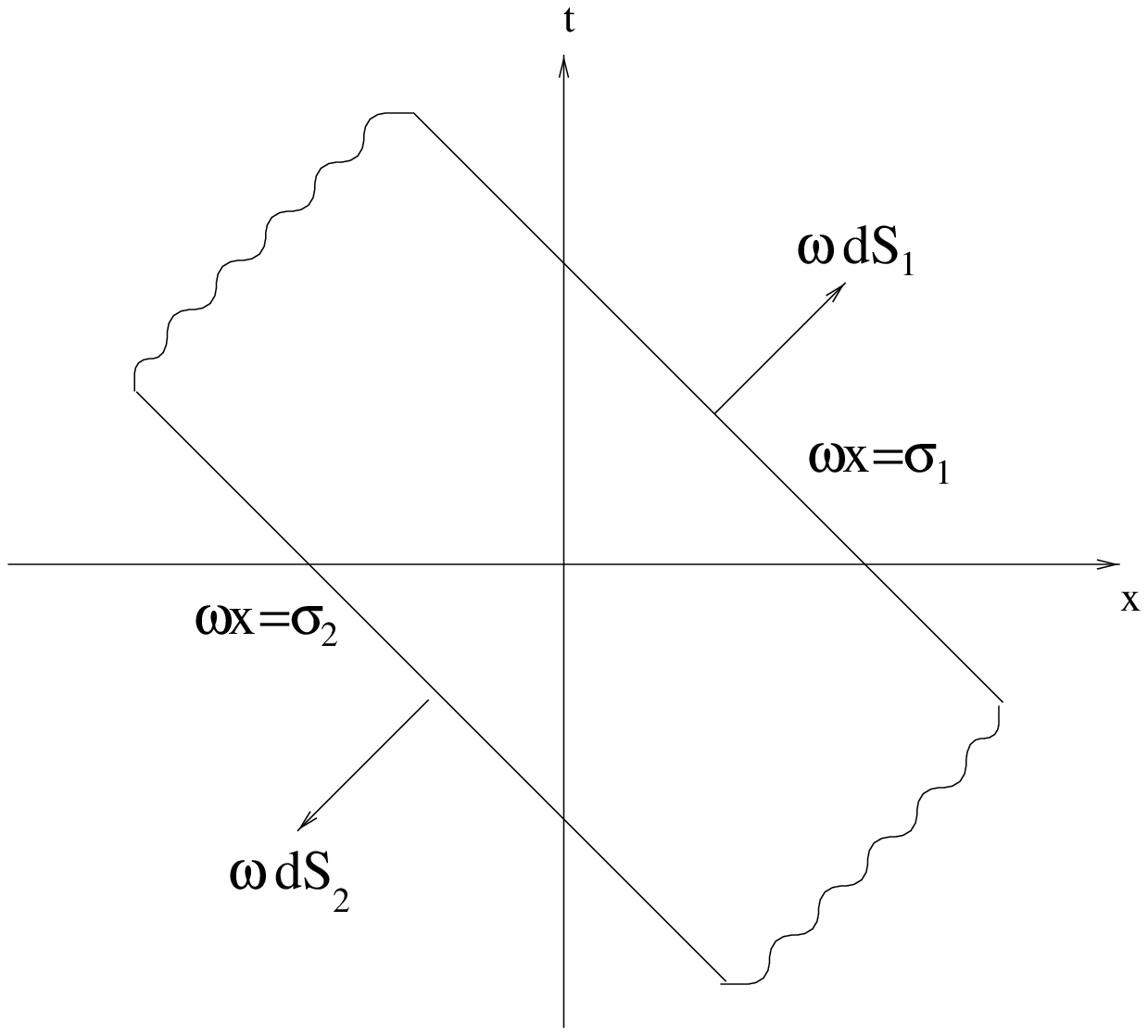}}
\caption{Flux in front-form of dynamics}\label{flux_LFD}
\end{minipage}
\end{center}
\end{figure}

Equation (\ref{FN}) leads to a conservation law only if $V$ is such that "lateral flux" vanishes.
In the usual approach of dynamics --
 denoted "instant-form" in Dirac classification \cite{Dirac_RMP_49} --
$V$ is the space-time volume limited by the hyperplane t=cte (see figure \ref{flux_xt}).
In Light-Front Dynamics -- denoted "front-form" in \cite{Dirac_RMP_49} --
$V$ is limited by two space-time planes $\omega\cdot x=\sigma$ (see figure \ref{flux_LFD}).

Let us illustrate the procedure in case transformation (\ref{Lorentz}) is a translation
\[ x_{\nu}\rightarrow x_{\nu}'= x_{\nu} + a_{\nu}\]
Noether theorem provides a set of four 4-currents
$\left\{T^{\mu\nu}\right\}_{\nu=0,3}$  given by
\[T^{\mu\nu}={\dLdmuphi}{\dNuphi}-g^{\mu\nu}{\cal L}\]
and satisfying
\[ \partial_{\mu}T^{\mu\nu}(x)=0  \]
They constitute the energy-momentum tensor.
Integrating the continuity equation over an arbitrary space-time  $V$ one gets
\[ \int_V d^4x\; \partial_{\mu}\; T^{\mu\nu}= 0\]
and choosing $V$ like in figure \ref{flux_xt} gives
\[ \left[\int d\vec{x}
T^{0\nu}(x) \right]_{upper} + \left[\int d\vec{x} T^{0\nu}(x)\right]_{lower}=0\]
That means the four quantities
\[ \hat{P}^{\nu}(t_0)=\int d\vec{x} T^{0\nu}(x) \]
are independent of the hyperplane $t=t_0$ position.

Alternatively, by choosing space-time planes $\omega\cdot x=\sigma$,
one obtains the $\omega$-dependent generators
\begin{equation}\label{GLGD}
\fbox{$\displaystyle \hat{P}^{\mu}(\sigma)=\int
T^{\mu\nu}(x)\delta(\omega\cdot x-\sigma) \;\omega_{\nu}d^4x $}
\end{equation}
with
\[ \partial_{\sigma}\hat{P}^{\mu}(\sigma) =0 \]
Equation (\ref{GLGD}) is the starting point of the Light-Front Dynamics.

In the instant-form approach, the spatial components depend only on the non interacting
part of the lagrangian and are called
kinematical. The zero-th component is the interaction Hamiltonian and governs the dynamics
\begin{eqnarray*}
P^0     &=& H_{int} \cr
\vec{P} &=& \vec{P}_0
\end{eqnarray*}
In LFD, all generators split into a free part ($\hat{P}^{\mu}_0$) and an interaction-dependent  one
($\hat{P}^{\mu}_{int}$)
\[\fbox{$\displaystyle  \hat{P}^{\mu}(\sigma)  = \hat{P}^{\mu}_0+ \hat{P}^{\mu}_{int}(\sigma) $}\]
with the interaction dependent part being along the $\omega$-direction
\[\fbox{$\displaystyle \hat{P}^{\mu}_{int}(\sigma)=\omega^{\mu}\;\int H_{int}(x)\delta(\omega\cdot x-\sigma)d^4x$}\]
This can be turn to profit to obtain a maximum number of seven non interacting
generators, an advantage that was already noticed by Dirac in his seminal
paper \cite{Dirac_RMP_49} about the different forms of relativistic dynamics.

Once obtained the generators, (\ref{P2})
provides the dynamical equation determining the mass of the system $M^2$.
After some algebra, one is led to
\begin{equation}\label{eqlfd0}
\fbox{$\displaystyle (M^2 - P_{0}^{2}) \mid\Psi> = 2 P_{0}\cdot\omega \; \int \;
{\cal{H}}_{int}(\omega \tau) \exp(-i \sigma \tau) \; d \tau \mid\Psi>  $}
\end{equation}
where ${\cal{H}}_{int}(k)$ denotes the Fourier transform of the hamiltonian density
\begin{eqnarray*}
{\cal{H}}_{int}(k)=\int {\cal{H}}_{int}(x) \exp(ik\cd x) d^4x
\end{eqnarray*}
We will consider hereafter the $\sigma$-stationary solutions and
set $\sigma=0$ in (\ref{eqlfd0}). The evolution of states towards $\sigma$ is
given in \cite{MMB_These_01}.

The state vector is decomposed into its Fock components $\Psi_{\alpha\beta}$
\footnote{We consider two interacting fields with corresponding  creation operators
$a^{\dagger}_{k}$ and $b^{\dagger}_{q}$  }
\begin{eqnarray}
\mid\Psi> &=& \sum_{\alpha\beta}\int d^4k_1\ldots d^4k_{\alpha}d^4q_1\ldots d^4q_{\beta} \cr
&&
\Psi_{\alpha\beta}(k_1,\ldots,k_{\alpha},q_1\ldots q_{\beta})  \;
   a^{\dagger}_{k_1} \ldots a^{\dagger}_{k_\alpha} \;
   b^{\dagger}_{q_1} \ldots b^{\dagger}_{q_\beta}  \;\mid0>    \label{WF}
\end{eqnarray}
which are the wavefunctions. We can consider the set
$\Psi_{n_{\alpha\beta}}\equiv\{\Psi_{\alpha\beta}\}$ as the
components of an infinite dimensional vector $\Psi= (\Psi_1,
\Psi_2, \Psi_3, \dots)$, coupled to each other via the interaction
operator $\int \; {\cal{H}}_{int}(\omega \tau) \; d \tau$.
Equation (\ref{eqlfd0}) is thus an infinite system of coupled channels.
\begin{eqnarray*} (M^2-P_0^2) \pmatrix{\ldots\cr \Psi_2 \cr \Psi_3\cr \ldots } =
2 P_{0}\cdot\omega \;  \int \; {\cal{H}}_{int}(\omega \tau) d \tau
\pmatrix{ \ldots \cr\Psi_2 \cr \Psi_3 \cr \ldots}
\end{eqnarray*}
If we restrict ourselves to the two-
($\Psi_{2}\equiv\{\Psi_{20}\}$)  and three-body
($\Psi_{3}\equiv\{\Psi_{21}\}$) wavefunctions, we obtain a system
of two coupled equations for $\Psi_2$ and $\Psi_3$ which
constitutes the ladder approximation. By expressing $\Psi_3$ in
terms of $\Psi_2$, one gets an integral equation for $\Psi_2$ with energy dependent kernel.

The construction of non zero angular momentum states
in LFD is made difficult by the fact
that generators of spatial rotations
contain the interaction. Consequently
the eigenvalue equation for $\hat{S}^2$ and $\hat{S}_3$ are non trivial
dynamical equations with a level of complexity higher than the mass equation (\ref{eqlfd0}) one.
To circumvent this difficulty we have made use of a kinematical angular momentum operator
\[ \vec{J}=-i\left(\vec{k}\times{\partial\over\partial{\vec{k}}}
          +        \hat{n}\times{\partial\over\partial{\hat{n}}}\right)\]
whose eigenstates can be constructed by using the standard angular momentum algebra.
The physical solutions are then obtained as a linear combination
of the kinematical operator $\hat{A}^2\equiv(\vec{J}\cdot\hat{n})^2$ eigenstates,
satisfying the so called "angular condition".
This method was shown to restore -- at least in a simple model -
the rotational invariance of the theory to a high degree of accuracy \cite{KCM_LCM_01}.
Its explanation requires long technical developments and will not be detailed here.
The interested reader could find the first published results in
\cite{KCM_Taiw_01,MCK_Heid_00,KCM_LCM_01}.

An important property of Light-Front Dynamics is the fact that the vacuum of the theory
is decoupled from any other Fock state despite the interaction.
This "emptiness of the vacuum" results into the absence of vacuum
fluctuations diagrams and considerably simplifies the evaluation of certain processes.

An appreciable advantage of this formalism with respect to other relativistic approaches,
is the clear link with the usual non relativistic dynamics: LFD wavefunctions
have the same physical meaning of probability amplitudes than their non relativistic counterparts.

Among the drawbacks we note --
apart from the psychological barrier of using a non conventional formalism --
the appearance of contact (instantaneous) interactions for non scalar constituents
and two direct consequences of solving the dynamical equations in a truncated Fock space:
the $\omega$ dependence  of the on-shell approximate amplitudes
and the fact that the different projections
of the non-zero angular momentum states along the $\vec{\omega}$
direction are non degenerate (the so called violation of the rotational invariance).

It could be of some help to say  few words about the words in what concerns
the Light-Front world.
Historically it appeared first in Dirac classification with the particular
choice of front surface $t-z=0$.
It reappeared in Quantum Field Theory in 1966,
under the label "Infinite Momentum Frame" calculations,
when it was  realized \cite{WEINBERG_66} that boosting the frame in which 
calculations of invariant amplitudes are made, gave easier results.
In particular all diagrams related to vacuum fluctuations automatically vanished.
Later on, it was shown \cite{CM_69}  that this procedure was equivalent to use the LFD
variables $z\pm t$ in the theory from the very beginning
and constitutes now the so called standard LFD, developped by many authors
\cite{Coester,Glazek,Ji,Fuda,Burkart,Brodsky,Bakker,Frederico,Hiller,Miller}.

The difference with respect to our approach is in the words "Explicitly"
and "Covariant" which we will try to justify.
To choose a front surface in the form $\omega\cdot x=\sigma$ is not only a mathematical
"delicatesse" but a way to carry everywhere in the theory the $\omega$-dependence
of the formal objects in an explicit way.
It has several advantages, all related to the fact that $\omega$ is a four vector
with well defined transformation properties.
For instance it provides explicitly covariant expressions for the on shell amplitudes,
a property  which is often hidden in the standard LFD formulation.
The latter one is recovered by fixing the value $\omega=(1,0,0,-1)$
but this value is associated to a particular reference frame
and it is not valid in any other one, what implies a non covariant formulation.
Another advantage of the covariant formulation is the fact that,
because the $\omega$-dependence is explicit, it can be controlled at
will and removed when calculating observables (e.g. in the form factors \cite{KS_94}).

If by covariance of a theory, we understand the fact that is based on
a set of generators satisfying the Poincar\'e algebra, then the standard and the explicitly covariant
approaches would be both covariant if they were solved in their full complexity.
This is never the case in practice, in particular due to the Fock space truncation, and
both formulations can lose this property when approximate schemes are used.

\section{Two Scalar particles}\label{Scalars}

It is always instructive to start by considering  a simple model
which could provide us with some physical insight with a lower
formal cost. We have thus considered the Wick-Cutkosky (WC) model
\cite{WC_54} describing two scalar particles ($m$) interacting
by a scalar exchange ($\mu$) with a  Lagrangian density ${\cal L}_I=g\varphi^2\chi$.
We are interested in bound states of two particles
with  momenta $k_1$ and $k_2$. In the reference frame
$\vec{k_1}+\vec{k_2}=0$ the ECLFD equation reads
\begin{equation}\label{LFD_WC}
 [4(k^2 +m^2)-M^2]
\Psi(\vec{k},\hat{n})=-{m^2 \over2\pi^3}\int{d^3k'\over\epsilon_{k'}}V(\vec{k},\vec{k'},\hat{n},M^2)
\Psi(\vec{k'},\hat{n})
\end{equation}
where $\vec{k}$=$\vec{k}_1$,
$\epsilon_{k}=\sqrt{k^{2}+m^2}$, $\omega=(1,\hat{n})$ in this particular frame
with $\hat{n}^2=1$  --
and $M$ the total mass of the system, related to its binding energy $B$ by $M=2m-B$.

The interaction kernel  is\footnote{Strictly speaking, the Wick-Cutkosky
model corresponds to the $\mu=0$ case. We consider here a trivial extension}
\[ V(\vec{k},\vec{k'},\hat{n},M^2)=-{4 \pi \alpha \over Q^2+\mu^2} \]
with
\begin{equation}\label{Q2}
Q^2=(\vec{k}-\vec{k'})^2
-(\hat{n}\cd\vec{k})(\hat{n}\cd\vec{k'}){(\epsilon_{k'}- \epsilon_{k})^2 \over\epsilon_{k'}\epsilon_{k}}
+\left(\epsilon_{k}^2+\epsilon_{k'}^2 -{M^2 \over 2}\right)
\left|{\hat{n}\cd\vec{k'}\over\epsilon_{k'}}-{\hat{n}\cd\vec{k} \over \epsilon_{k}}\right|
\end{equation}
Relativistic effects
are all included in the second and third $\hat{n}$-dependent terms of $Q^2$.
By formally setting  $\hat{n}=0$ in this expression
 -- though $\hat{n}^2=1$ ! --  one recovers the non relativistic Coulomb kernel.
By furthermore assuming $B<<m$, the kinematical left hand side term of (\ref{LFD_WC}) becomes
$4m(k^2+B)$ and the ECLFD equation coincides with the non relativistic Schrodinger
equation in momentum space for Coulomb
interaction\footnote{We note a missprint in page 465 of  \cite{MC_Evora_01}
where the expression $4mk^2+2mB$ was instead written}.

Equation (\ref{LFD_WC}) was solved  \cite{MC_PLB_00,MMB_These_01} 
for several values of $J$ and $\mu$.
For S- waves, the solution $\Psi$ is a scalar function depending on two scalar arguments
$\Psi=\Psi(k,\hat{n}\cdot\vec{k})$, i.e. the LFD S-wave wavefunction has an angular
dependence!
For non-zero angular momentum states the solution was obtained on a basis of
eigenfunctions of the $\hat{A}^2$ operator, discussed in the previous section.
In a truncated Fock space, 
these eigenfunctions were found to have different $M^2$ values \cite{KCM_Taiw_01,MCK_Heid_00} --
a result also noticed in \cite{CMP_PRC61_00} --
and a method was proposed to restore in the two-body sector the rotational invariance.
Our main results concerning S-waves are summarized in what follows.
The numerical values given hereafter all correspond to $\hbar=c=m=1$ units.

\paragraph{Size and nature of relativistic effects}

$ $
\bigskip

\begin{figure}[htbp]
\begin{minipage}[t]{83mm}
\epsfxsize=8.cm{\epsfbox{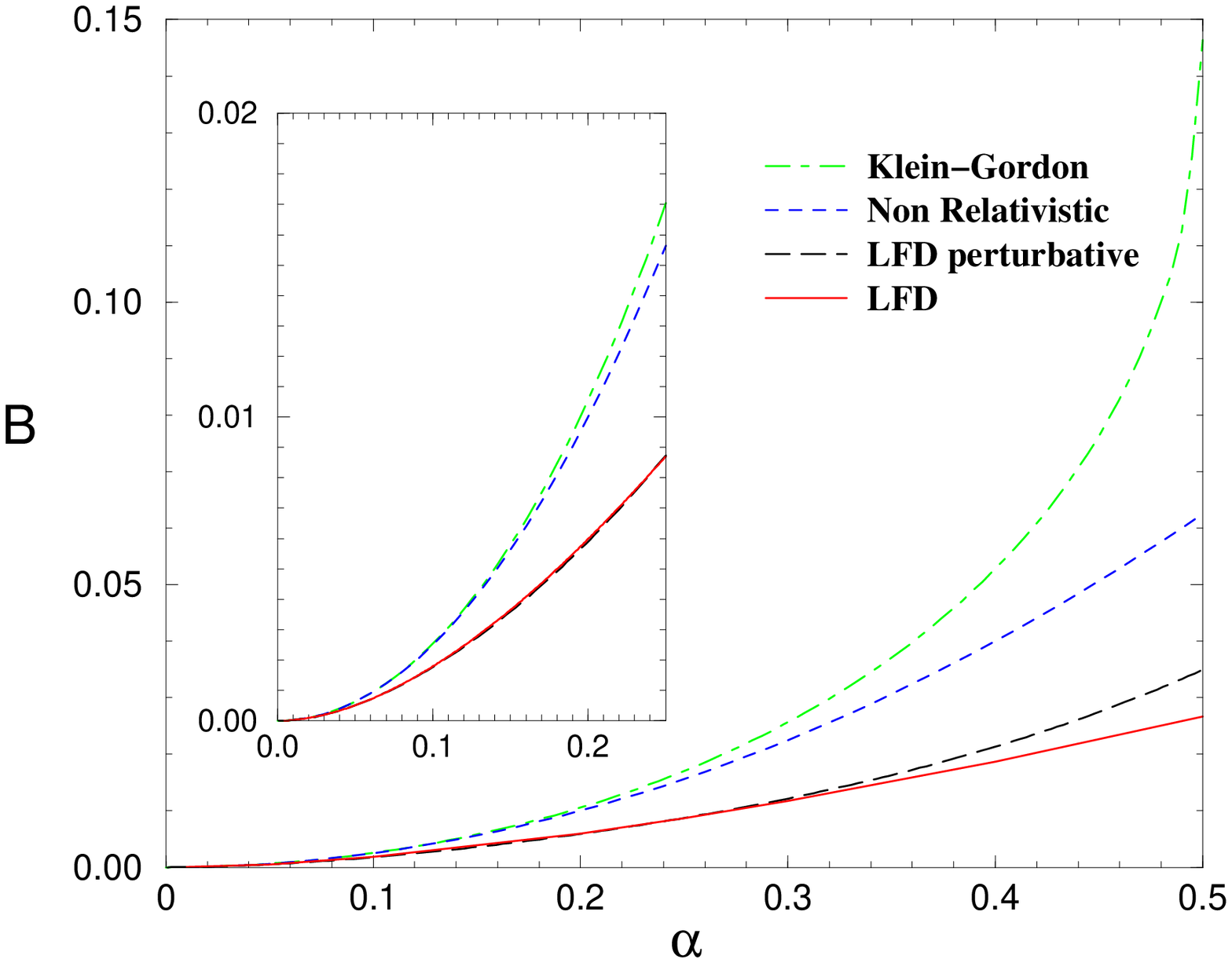}}
\caption{LFD energies (solid) for $\mu=0$ as a function of $\alpha$,
compared with non relativistic, Klein-Gordon and perturbative results}\label{NR_KG_LFDp}
\end{minipage}
\hspace{0.2cm}
\begin{minipage}[t]{83mm}
\epsfxsize=8.cm
{\epsfbox{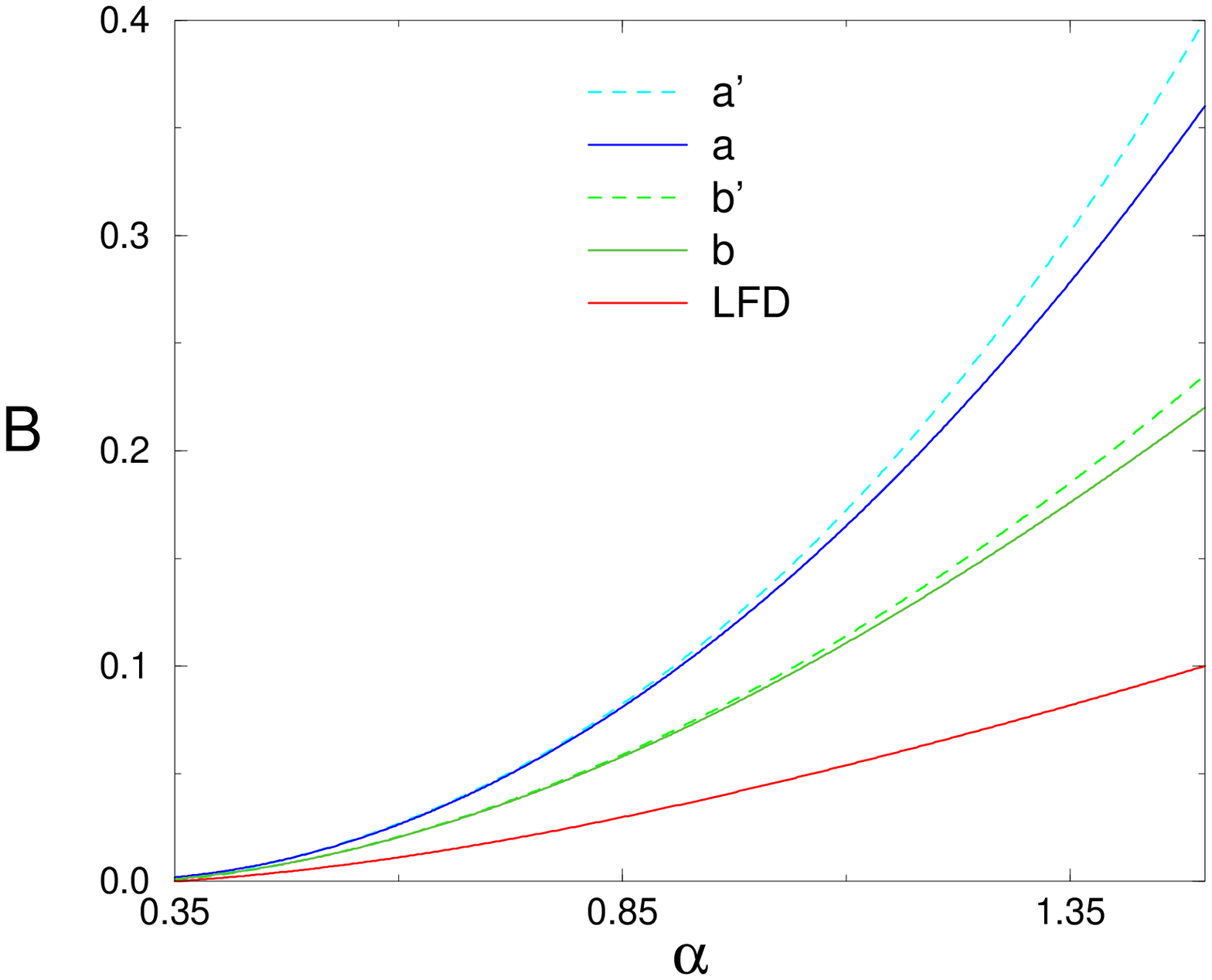}}
\caption{Contribution of different relativistic corrections (see text for details)}\label{Corrections}
\end{minipage}
\end{figure}

The size and nature of relativistic effects in the binding energies can be seen
in figure \ref{NR_KG_LFDp}.
The LFD results for $\mu=0$ are compared to the non relativistic ones
and to those given by the Klein-Gordon equation.
One can see a quick departure from Schrodinger results and a
clear repulsive character - smaller binding energy -
contrary to  Klein-Gordon (and Dirac) equations which produces attractive corrections to all orders.
LFD energies are also compared to a first order perturbative calculations 
-- equally valid for Bethe-Salpeter (BS) equation -- which have the form \cite{FFT_73}
\begin{equation}\label{bpert}
B_{pert}={m\alpha^2 \over 4}\left(1-{4 \over \pi }\alpha \log{1\over \alpha}\right)
\end{equation}
and turn to be relevant until  $\alpha\approx0.3$.

It is worth noticing that sizeable relativistic effects
are observed in a system for which both the binding energy and the average momentum are small.
At $\alpha=0.3$, for instance,
they account for a 100\%  effect in the binding
energy whereas  ${<k^2>\over m^2} \approx 2\%$.

We would also like to notice that the large differences between
the ECLFD and the non relativistic solutions are not of kinematical origin.
In order to disentangle the different contributions to the relativistic energies $B$,
equation (\ref{LFD_WC}) has been formally written  $K\Psi=\frac{1}{\varepsilon} V\Psi$
and we have considered several approximations of it.
The results are displayed in figure \ref{Corrections}.
We have first considered the case
of a non relativistic kernel V -- i.e. put $\hat{n}=0$ in
(\ref{Q2}) and $\epsilon=m$ -- together with
in curve {\it a} the non relativistic kinematics $K=4m(k^2+B)$
and in curve {\it a'} the relativistic one $K=4(k^2+m^2)-M^2$.
Curves {\it b} and {\it b'} are obtained in the same manner
but putting $\epsilon=\sqrt{k^{2}+m^2}$.
The last one corresponds to the full ECLFD equation.
These results show that -- at least in what scalars are concerned --
the main effect of a relativistic formalism
is not in the kinetic energy but in the interaction kernel.
The kinematical term $K$
has a very little influence on $B$, which furthermore goes on the opposite direction, whereas
$\epsilon$ and $V$ kernel contribution are both essential.
One can thus conclude that kinematical corrections
alone are not representative of relativistic effects.
Even when they are included in the kernel, e.g.
through $\epsilon_k$, they can only account for half the effect.

\paragraph{Comparison with Bethe-Salpeter equation}

$ $

\begin{figure}[htbp]
\begin{minipage}[htbp]{83mm}
\begin{center}
\mbox{\epsfxsize=8.cm\epsffile{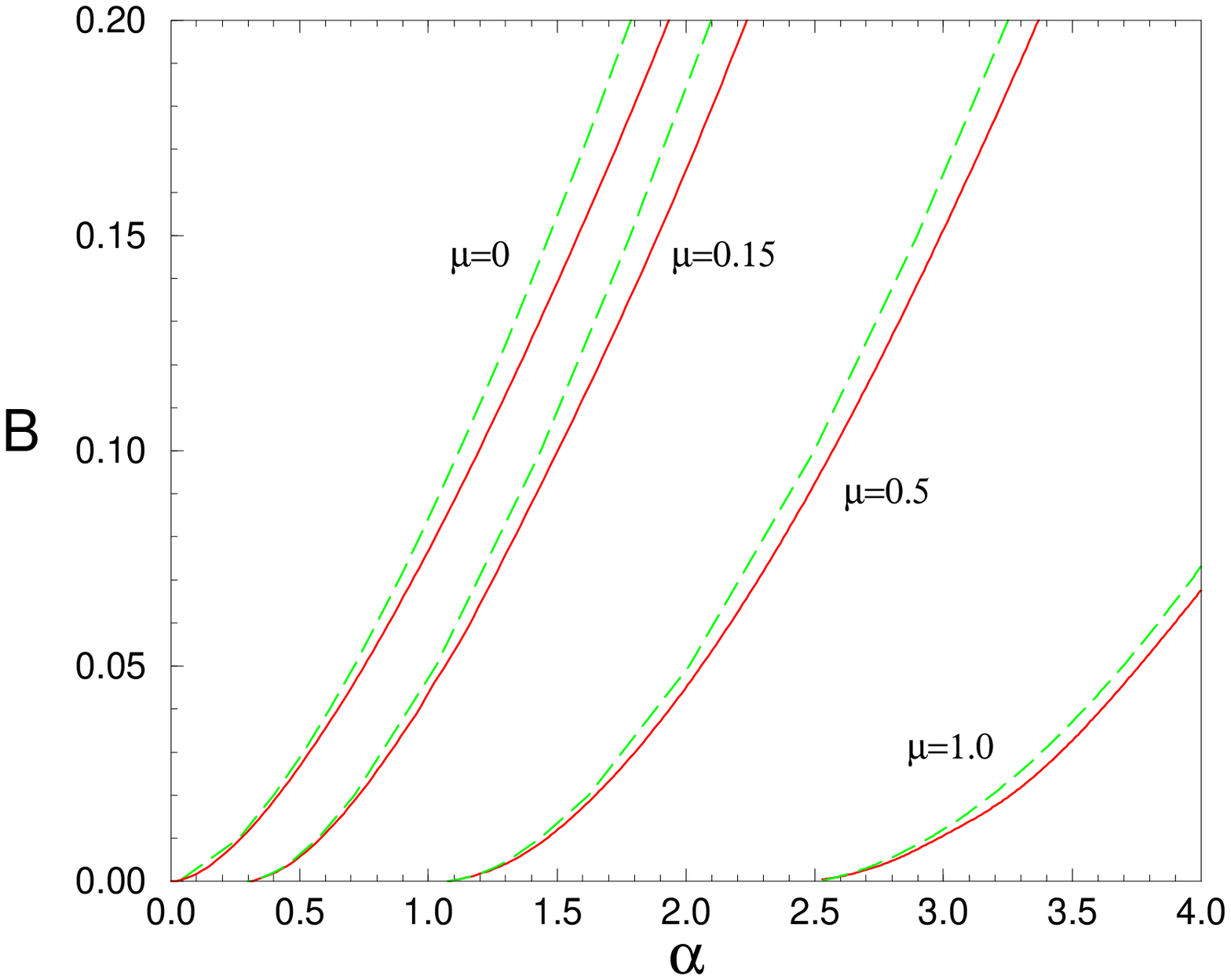}}
\caption{$B(\alpha)$ for different
values of $\mu$ in LFD (solid) and BS (dashed) approaches}\label{B_LFD_BS_mu}
\end{center}
\end{minipage}
\hspace{0.2cm}
\begin{minipage}[htbp]{83mm}
\begin{center}
\mbox{\epsfxsize=8.cm\epsffile{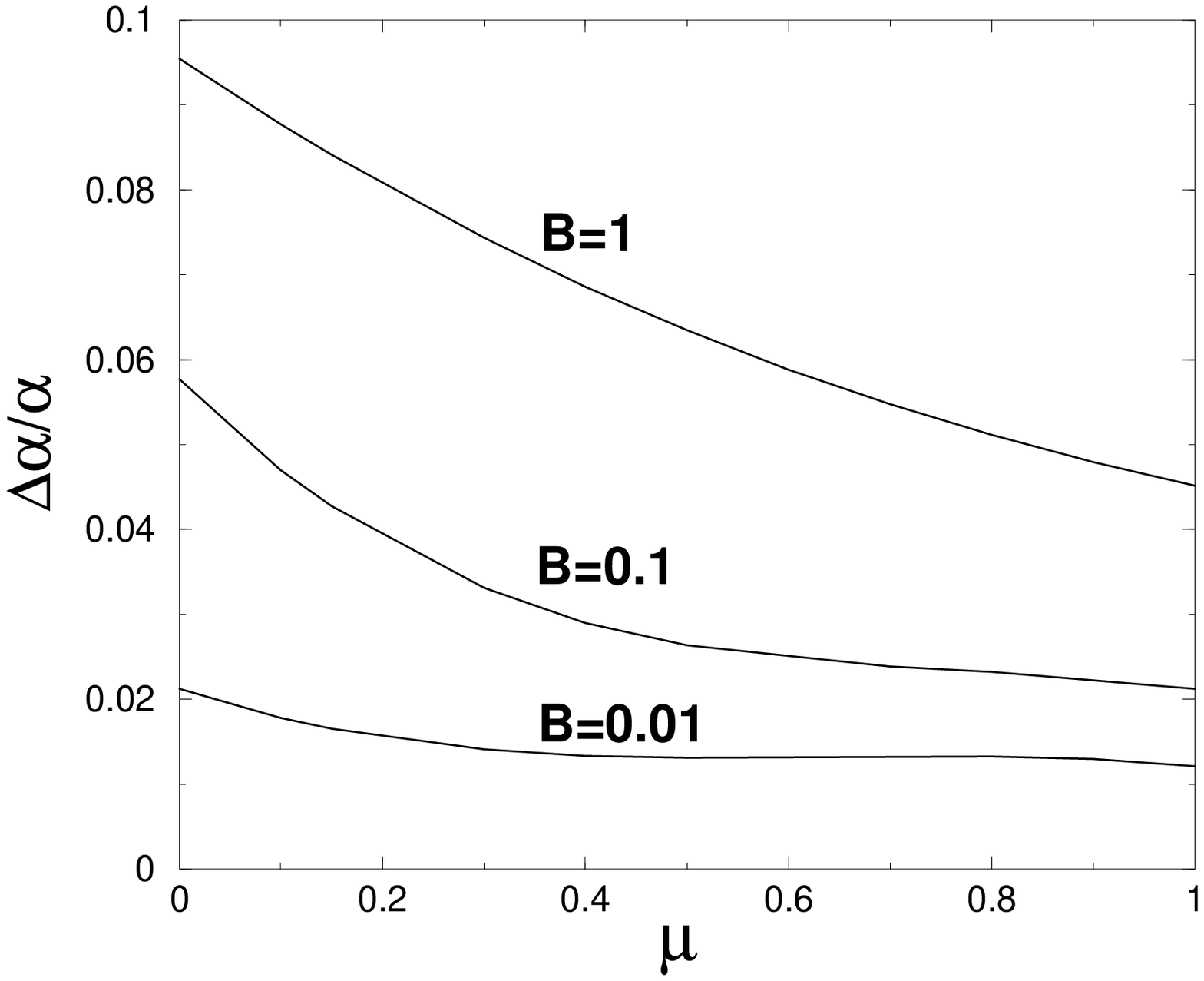}}
\caption{Differences in the coupling constant as a function of $\mu$
for  fixed values B.}\label{Dalpha_mu}
\end{center}
\end{minipage}
\end{figure}

\bigskip
The comparison with Bethe-Salpeter  equation is done in figures \ref{B_LFD_BS_mu} and \ref{Dalpha_mu}.
Figure \ref{B_LFD_BS_mu} represents the LFD  $B(\alpha)$ curves (solid line)
for different values of $\mu$.
They are compared with those provided by BS equation (dashed-line)
in the same ladder approximation,
whose kernel incorporates higher order intermediate states.
Their results are seen to be close to each other.
This fact is far from being obvious --
specially for large values of coupling constant -- due to the differences in their ladder kernel.
A quantitative estimation of their spread can be given by looking into an horizontal
cut of figure \ref{B_LFD_BS_mu}, i.e. calculating the relative difference in the coupling constant
$(\alpha_{LFD}-\alpha_{BS})/\alpha_{LFD}$ for a fixed value of the binding energy.
The results, displayed in figure \ref{Dalpha_mu} for $B=1.0,0.1,0.01$, show that relative differences
({\it i})  are decreasing functions of $\mu$ for all values of $B$
({\it ii}) increase with $B$ but are limited to 10\% for the strong binding case $B=m$
which involves values of $\alpha\ge 5$.
This indicates the relatively weak importance of including higher Fock components
in the ladder kernel even for strong couplings, as was noticed in \cite{Bakker}.

\paragraph{Weak binding limit and non relativistic solutions}

$ $

\bigskip

\begin{figure}[htbp]
\begin{minipage}[htbp]{81mm}
\begin{center}
\mbox{\epsfxsize=8.cm\epsffile{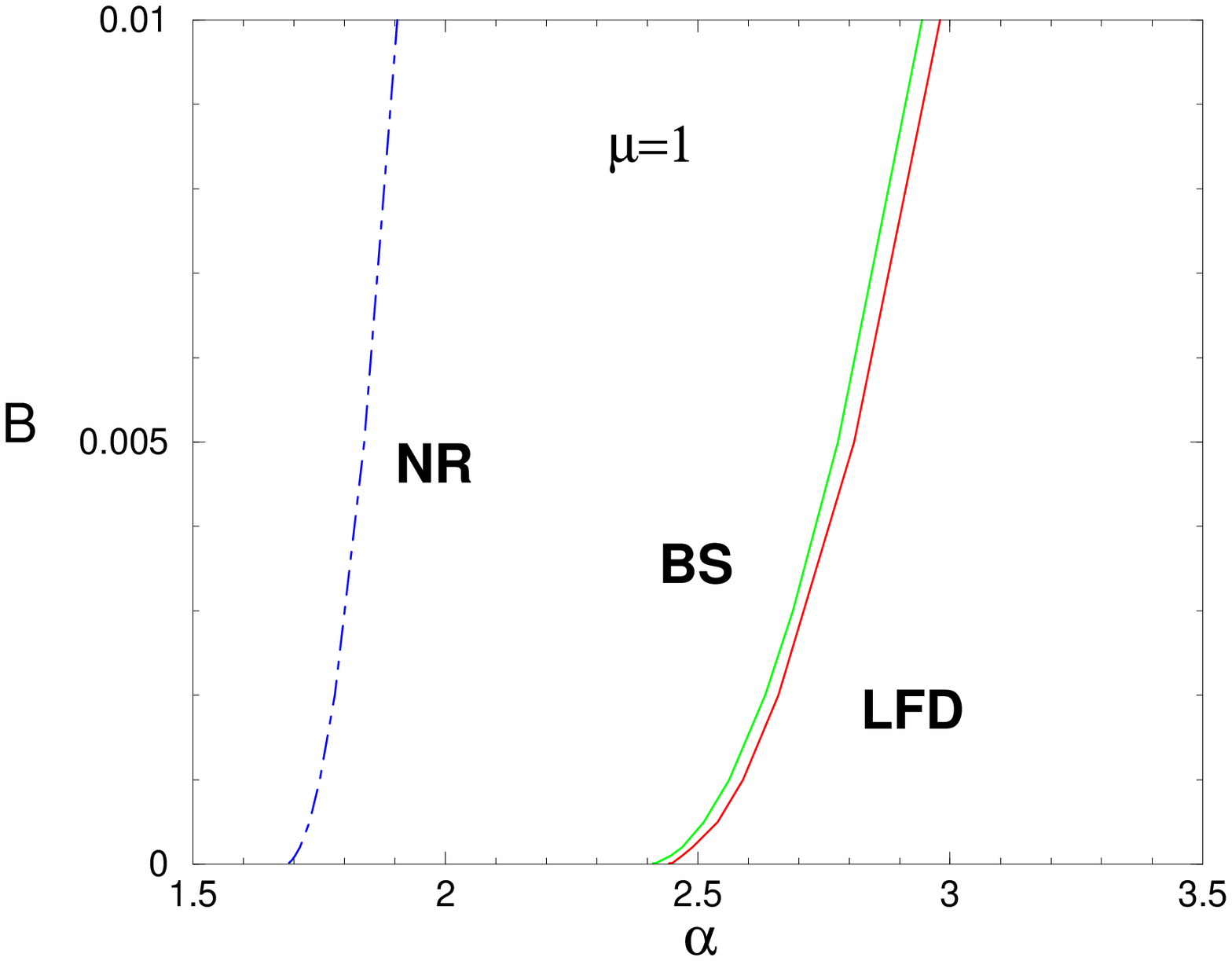}}
\caption{Zero binding energy limit of LFD and BS equations (solid)
compared with non relativistic solutions (dot-dashed) for $\mu=1$}\label{B_LFD_BS_NR_mu=1}
\end{center}
\end{minipage}
\hspace{0.5cm}
\begin{minipage}[htbp]{81mm}
\begin{center}
\mbox{\epsfxsize=8.cm\epsffile{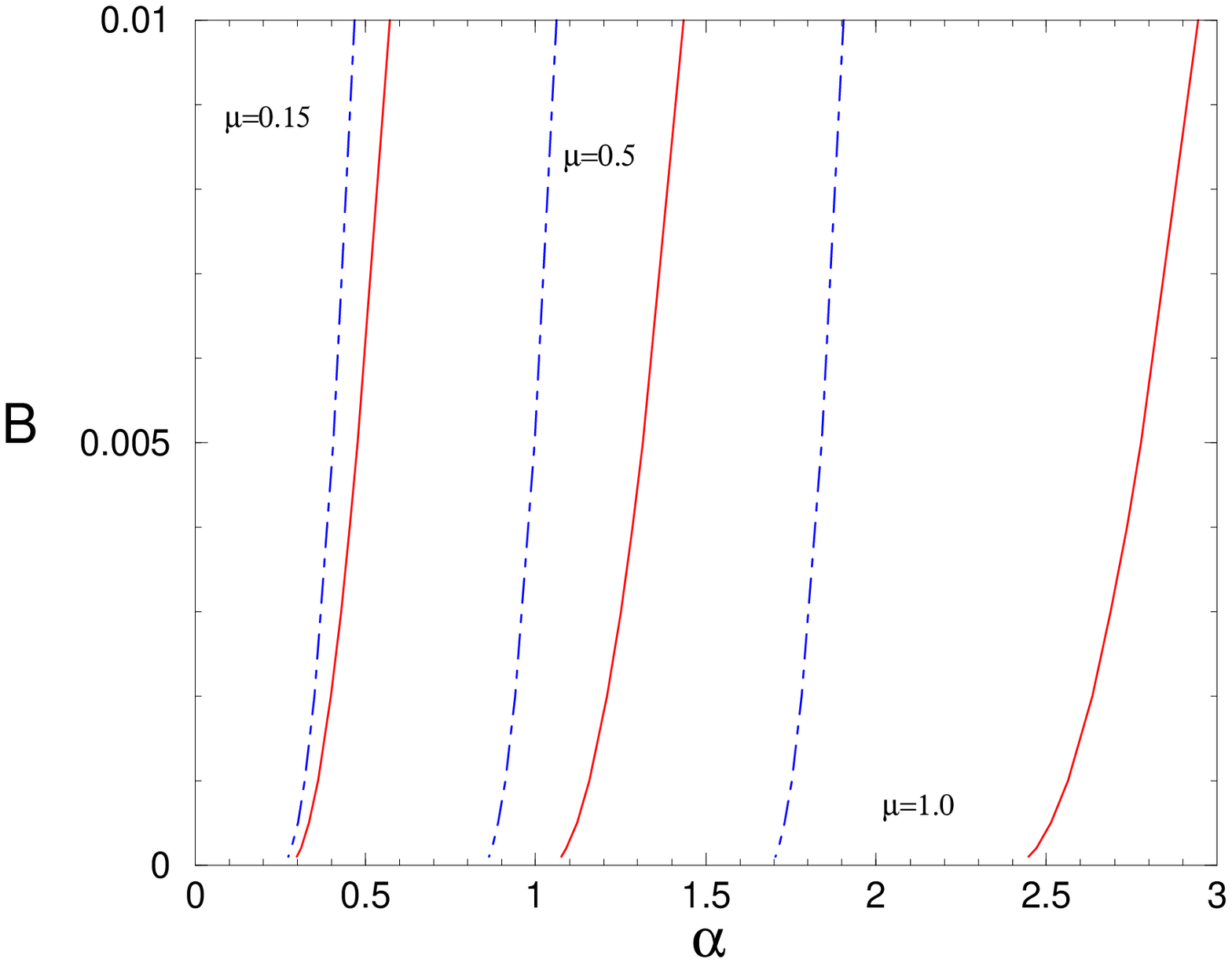}}
\caption{The same study for different values of $\mu$. LFD and BS curves
are not distinguished}\label{B_NR_LFD_BS_mu}
\end{center}
\end{minipage}
\end{figure}

An interesting result was obtained by studying the weak binding limit of
B($\alpha$). We found that, except for the case of zero mass
exchange, relativistic and non relativistic solutions
differ even when describing zero binding energy systems.
These results are displayed in figures \ref{B_LFD_BS_NR_mu=1} and
\ref{B_NR_LFD_BS_mu}. BS solutions, also included, display the same behaviour, indicating that
this is not a pathology of the Light-Front formalism but seems rather a general
feature of consistent relativistic theories.
A system bound by a massive exchange would be described by different parameters whether one uses
LFD (and BS) or Schrodinger approaches, no matter how small the binding energy will be.
We concluded from that to the non existence of non relativistic limit for $\mu\ne0$.

We would like to mention here that an important part of these differences
were taken into account by including in  the Schrodinger equation
energy dependent interactions \cite{Desplanques}.

\paragraph{The scalar deuteron}

$ $

\bigskip

A straightforward application to this model, and a way
to evaluate the modifications induced by
a relativistic treatment of a semi-realistic model, was done by
building a scalar model for deuteron. By adding a "repulsive
scalar" term\footnote{One should note that such a repulsive interaction
cannot arise from a scalar exchange but can mimic some fermionic interactions}
one gets a relativistic version of the Nucleon-Nucleon MT potential \cite{MT_69}.
\begin{equation}\label{MT_LFD}
 V=V_R+V_A={\lambda_R\over Q^2+\mu_R^2}-{\lambda_A\over Q^2+\mu_A^2}
\end{equation}
In the non relativistic limit, $\lambda_R,\mu_R,\lambda_A,\mu_A$
are adjusted to reproduce a realistic deuteron wavefunction with
$B=2.23$ MeV and acceptable NN scattering parameters. If one
inserts kernel (\ref{MT_LFD}) in the LFD equations, the binding
energy is shifted from $B=2.23$ to $B=0.96$ MeV, a dramatic repulsive
effect. One can recover the physical value for B by shifting
$\lambda_R=7.29 \rightarrow\lambda_R=6.60$ -- all other parameters
being unchanged -- what makes a 10\% difference in a coupling constant.
The spectacular change in $B$ is partially due to the small value of
the deuteron binding energy. It illustrates however
well the difficulty to unambiguously determine a coupling constant
in the strong interaction physics, even for systems
widely  considered as being {\it a priori} non relativistic.

\paragraph{Wavefunctions}

$ $

\begin{figure}[htbp]
\begin{minipage}[htbp]{81mm}
\begin{center}
\mbox{\epsfxsize=8cm\epsffile{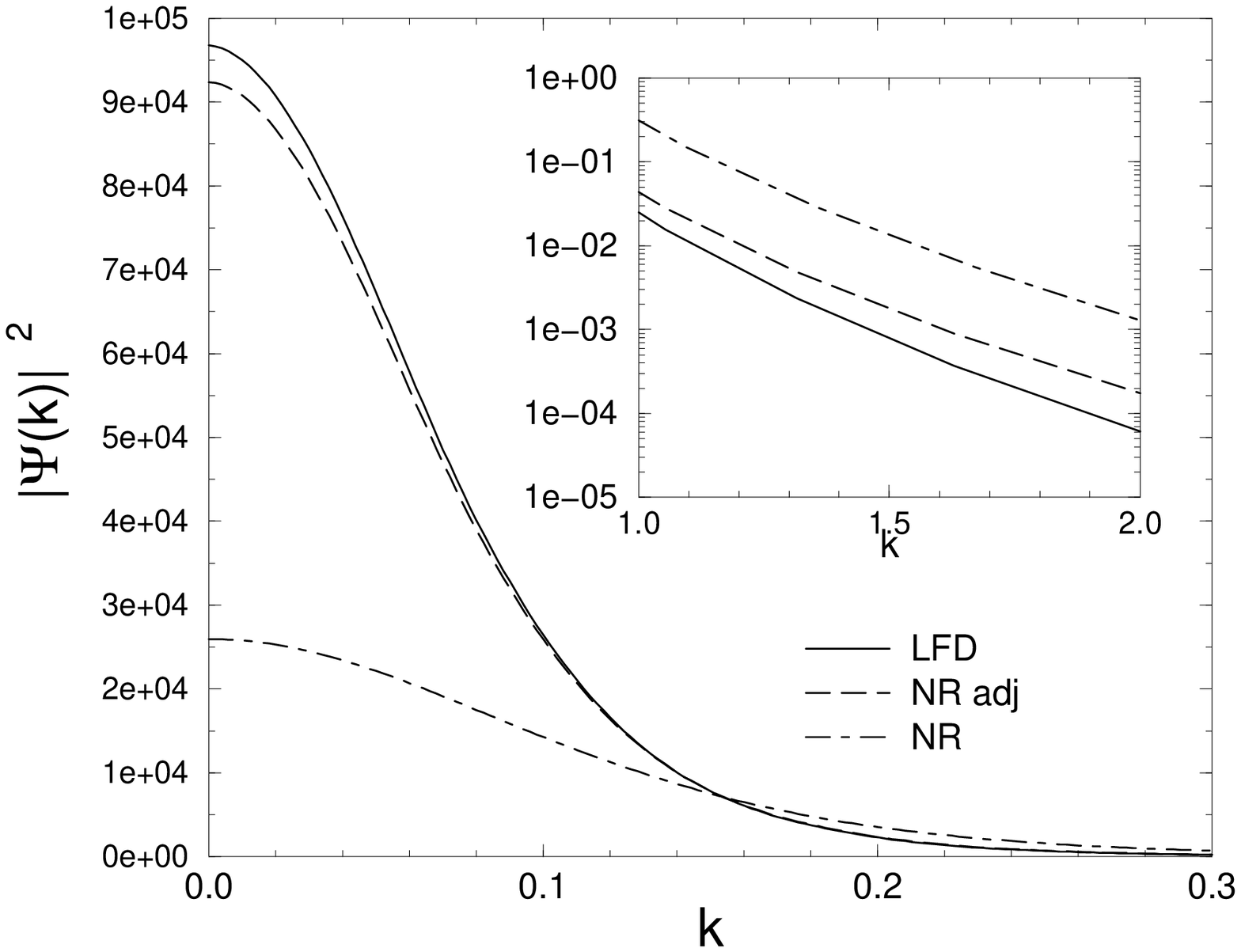}}
\caption{LFD and Schrodinger wavefunctions, squared and integrated
over $\theta$ for $\mu=0$.}\label{wf2_alpha=0.5_mu=0}
\end{center}
\end{minipage}
\hspace{0.5cm}
\begin{minipage}[htbp]{81mm}
\begin{center}
\mbox{\epsfxsize=8cm\epsffile{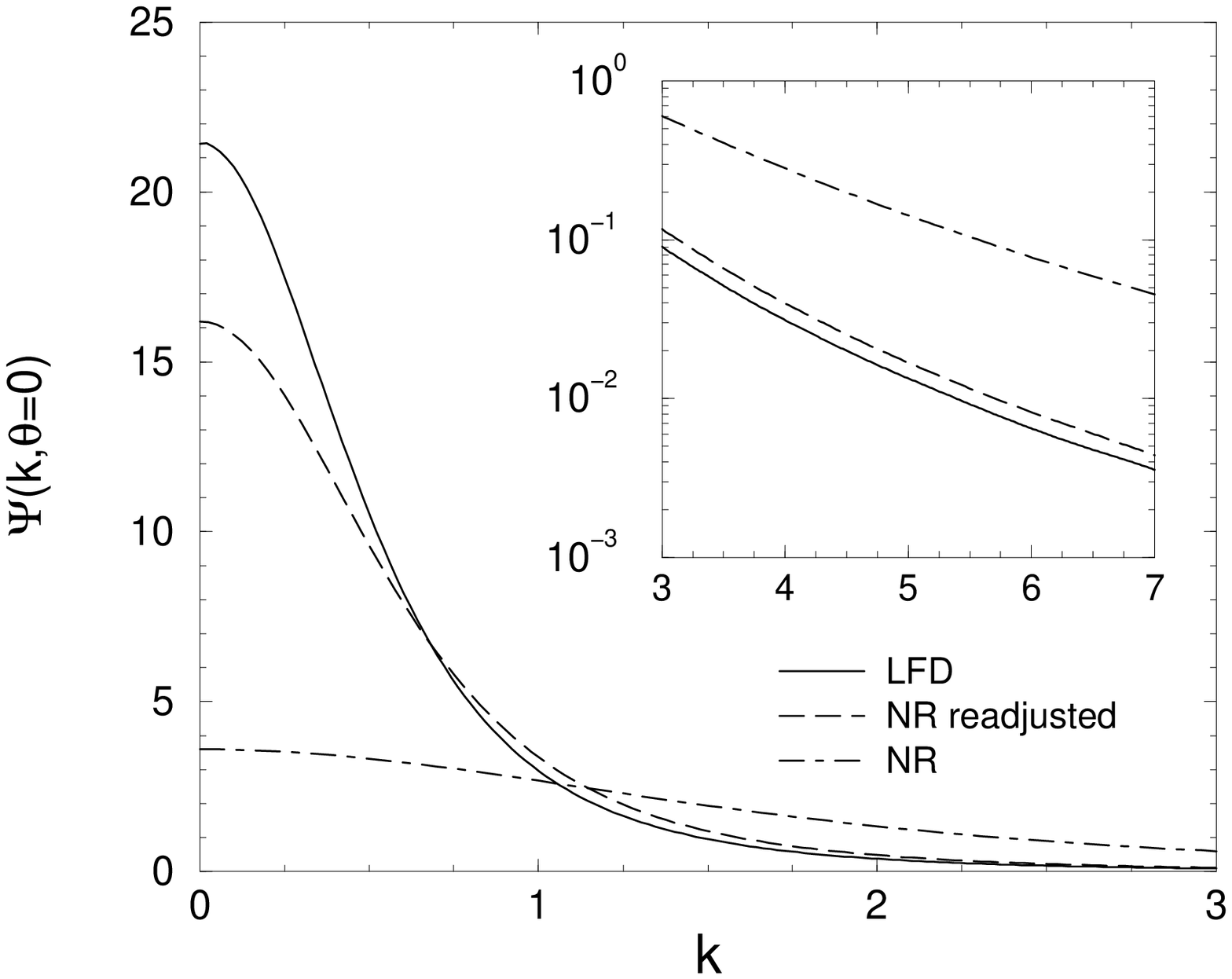}}
\caption{Similar comparaison in a highly relativistic case ($\alpha=5$)}\label{wf_alpha=5_mu=0}
\end{center}
\end{minipage}
\end{figure}

\bigskip
The main modification when comparing LFD and NR wavefunctions
with the same coupling constant
is due to the change in the corresponding binding energies.
A first requirement to compare them
is thus to re-adjust one of the $\alpha$ values (we choose the NR one)
in order to deal with states of equal $B$.
Figure \ref{wf2_alpha=0.5_mu=0} shows the comparison of these wavefunctions,
squared and integrated over the angular dependence $\hat{n}\cdot\hat{k}\equiv\cos\theta$,
for $\mu=0$ and a moderate coupling  $\alpha=0.5$.
One then has $B_{LFD}$=0.0267 and $B_{NR}$=0.0625
which has been re-scaled to the LFD value with $\alpha=0.327$.
One can see small deviations ($\approx$5\%) at $k=0$.
In the relativistic region, the LFD solutions
are systematically smaller than the NR ones and their differences increase
reaching a factor 3 for $k=2$, despite the moderate values of B and $\alpha$.
When dealing with highly relativistic systems -- like e.g. mesons in the
constituent quark model --
both descriptions differ even for small values of $k$.
Figure \ref{wf_alpha=5_mu=0} shows the case  $B_{LFD}$=0.84 and $\alpha=5$
where the squared wavefunctions at $k=0$ are $\approx$70\% different.

\section{Two fermions system}\label{Fermions}

We have also obtained \cite{MCK_PRD_BR_01,
MCK_PRD_01,MMB_These_01} recently the LFD bound state  solutions
for a system of two fermions
 interacting  with the usual -- scalar (S), pseudoscalar (PS), pseudovector (PV)
and vector (V) -- OBEP Lagrangians:
\begin{eqnarray*}
{\cal L}_s    &=& g_s \bar{\Psi} \phi \Psi \cr
{\cal L}_{ps} &=& g_{ps} \bar{\Psi} i\gamma_5\phi \Psi \cr
{\cal L}_{pv} &=& {g_{pv}\over2m} \bar{\Psi}\gamma_5\gamma_{\mu}\partial^{\mu}\phi \Psi \cr
{\cal L}_{v}  &=&\bar{\Psi}[g_{v}\gamma^{\mu}A_{\mu}\;+\;
{f_v\over2m}\sigma^{\mu\nu}(\partial_{\mu}A_{\nu}-\partial_{\nu}A_{\mu} )] \Psi
\end{eqnarray*}
As in the scalar case, calculations were performed in the ladder approximation.
The interaction kernel
is provided by the Born scattering amplitude ${\cal M}$ and -- according
to the Light-Front graph technique --  has two contributions
which differ from each other by the light-front time ordering.
\begin{figure}[hbtp]
\begin{center}
\begin{minipage}[t]{10mm} \vspace{-3.cm} ${\cal M}=$ \end{minipage}
\begin{minipage}[t]{70mm} \begin{center}\epsfxsize=60mm\mbox{\epsffile{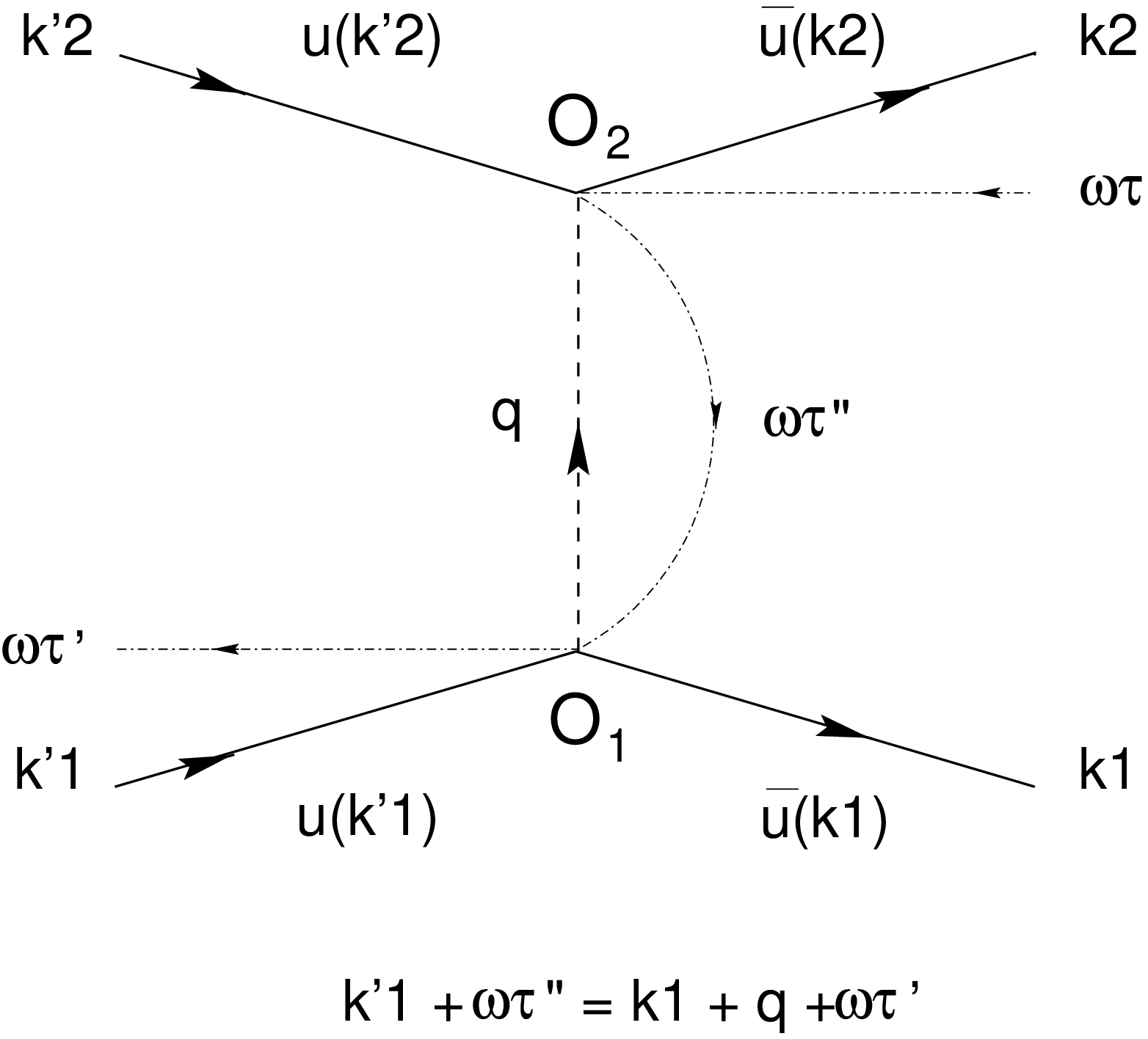}}\end{center}
\end{minipage}
\begin{minipage}[t]{5mm} \vspace{-3.cm}  {\bf +} \end{minipage}
\begin{minipage}[t]{70mm} \begin{center}\epsfxsize=60mm\mbox{\epsffile{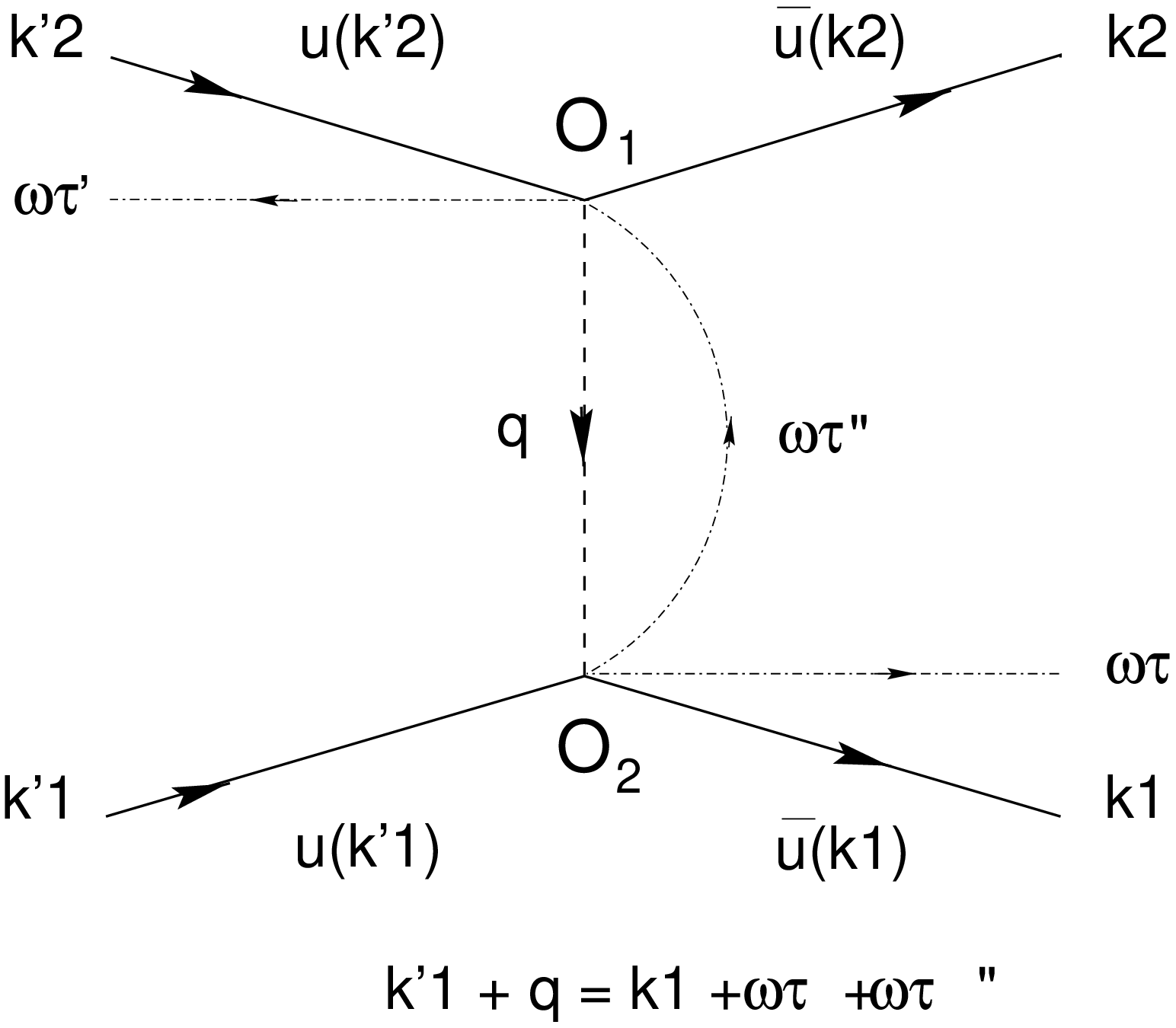}}\end{center}
\end{minipage}
\end{center}
\end{figure}
Its analytical structure has the form
\[ {\cal M}= \left\{ A_1(k'_1,k_1,\tau')+A_2(k_1,k'_1,\tau) \right\}
\left[ u_{\sigma'_1}(k'_1)\hat{O}_1 \bar{u}_{\sigma_1}(k_1) \right] \left[
u_{\sigma'_2}(k'_2)\hat{O}_2 \bar{u}_{\sigma_2}(k_2) \right] \]
where $\hat{O}_i$ are vertex operators depending on
the type of coupling ($\hat{O}_i={\bf 1},i\gamma_5,\ldots$)
and $A_1,A_2$ scalar functions.
In "center of mass" variables, i.e. boosting to a reference frame in which
$\vec{k}_1+\vec{k}_2=0$,  the scalar part can be written as
\[A_1(\vec{k},\vec{k}\,',\hat{n},M) + A_2(\vec{k},\vec{k}\,',\hat{n},M)
={1\over Q^2+\mu^2} \]
with $Q^2$ given by (\ref{Q2}) and the total amplitude reads
\[{\cal M}_{\sigma'_1\sigma'_2,\sigma_1\sigma_2} = {1\over Q^2+\mu^2}\;
\left[u_{\sigma'_1}(k'_1) \hat{O}_1 \bar{u}_{\sigma_1}(k_1)\right]\;
\left[u(k'_2)_{\sigma'_2} \hat{O}_2 \bar{u}(k_2)_{\sigma_2}\right] \]

As in the scalar case, the two-fermion LFD wavefunctions  $\Phi$
are Fock components of state vector (\ref{WF}). Their spin
structures depend on the quantum numbers of the state. For
$J^{\pi}=0^+$ they have the form
\begin{equation}\label{0PLUS_WF}
\Phi_{s_1s_2}(k_1,k_2,p,\omega)=
\bar{u}_{s_2}(k_2)\;\;\hat{\bf\phi}\;\;\gamma_5U_C\;\bar{u}_{s_1}(k_1)
\end{equation}
with
$$\hat{\bf\phi}=\phi_1+ \left[ {2m\gamma_{\mu}\omega^{\mu}\over\omega\cdot p}-
{4m^2\over s} \right]\phi_2$$
depending on two scalar functions  $\phi_i$.
For $J^{\pi}=1^+$ they read
\begin{equation}\label{1PLUS_WF}
\Phi^{\lambda}_{s_1s_2}(k_1,k_2,p,\omega)=\epsilon^{\lambda}_{\mu}(p)
\;\bar{u}_{s_2}(k_2)\;\hat{\bf\phi}^{\mu}\;U_C\;\bar{u}_{s_1}(k_1)
\end{equation}
in which
\begin{eqnarray*}
\hat{\bf\phi}_{\mu}&=&\phi_1{(k_1-k_2)_{\mu}\over2m^2}
                +     \phi_2{\gamma_{\mu}\over m}
                +     \phi_3{\omega_{\mu}\over\omega\cdot p}
                +     \phi_4{(k_1-k_2)_{\mu}\hat{\omega}\over2m(\omega\cdot p)}    \cr
               &+&    \phi_5{i\gamma_5\epsilon_{\mu\nu\rho\sigma} (k_1+k_2)_{\nu}(k_1-k_2)_{\rho}\omega_{\sigma}\over2m^2\omega\cdot p}
                +     \phi_6{m\omega_{\mu}\hat{\omega}\over(\omega\cdot p)^2}
\end{eqnarray*}
$p$ is the total momentum, $U_C=\gamma_2\gamma_0$ the  charge
conjugation operator,  $e^{\lambda}_{\mu}(p)$  the polarization vector and
$\hat{\omega}=\gamma_{\mu}\omega^{\mu}$.
They depend on six scalar fonctions  $\phi_i$.

To get easier links with the non relativistic wavefunctions one
uses instead of $\phi_i$ some linear combinations denoted $f_i$.
For $0^+$ states, for instance, one simply has
\[\begin{array}{lcll}
f_1  &=& 2\sqrt2\epsilon_k      &  \phi_1  \cr
f_2  &=& 2\sqrt2k {m\over\epsilon_k}&  \phi_2
\end{array}\]
with $f_1$ tending to the usual NR wavefunction and
$f_2$ being an extra component of relativistic origin with no counterpart.
For $1^+$ states the combination is more involved \cite{CK_NPA581_95}
but the $f_i$ functions have similar properties.
Thus $f_1$ tends to the non relativistic S-wave component $u_S$,
$f_2$ tends to the D-waves $-u_D$ and $f_{3-6}$ are extra components
of relativistic origin.

One can see that in this $f$-representation, LFD provides a very
interesting parametrization of the relativistic dynamics.
One deals with similar formal objects, depending on similar arguments.
In this way, relativity acts by modifying the usual
wavefunctions and by introducing extra components
which becomes negligible -- or tends to zero -- in the non relativistic regime.

In $\Phi$ representation (\ref{0PLUS_WF}-\ref{1PLUS_WF}) the two-body LFD equation reads
\begin{eqnarray} \left[M^2-(k_1+k_2)^2\right]
\Phi_{\sigma_1\sigma_2}^{\lambda}(k_1,k_2,p,\omega\tau) &=&{m^2\over2\pi^3} \sum_{\sigma'_1\sigma'_2}
\int{d^3k'_1\over2\epsilon_{k'_1}}{d^3k'_2\over2\epsilon_{k'_2}}d\tau'
\delta(k'_1+k'_2-p-\omega\tau)   \label{EQWF}\\ \nonumber \;2(\omega\cdot
p)\;\times\; \Phi_{\sigma'_1\sigma'_2}^{\lambda}(k'_1,k'_2,p,\omega\tau')& \times &
V_{\sigma'_1\sigma'_2,\sigma_1\sigma_2}(k'_1,k'_2,p,\omega\tau';k_1,k_2,p,\omega\tau)
\end{eqnarray}
Inserting (\ref{0PLUS_WF}-\ref{1PLUS_WF}) in (\ref{EQWF}) one is left -- after some lengthy algebra --
with a set of coupled two-dimensional integral equations on  the form
\begin{equation}\label{eqres}
A_{\alpha}(k) \phi_{\alpha}(k,u)
= \sum_{\beta} \int dk'du'\;{B}_{\alpha\beta}(k,u,k',u') \phi_{\beta}(k',u')
\end{equation}
where $u=\hat{n}\cdot\hat{k}$. $A_{\alpha}$ contains kinematical terms and
$B_{\alpha\beta}$ the interaction kernel.
For $J^{\pi}=0^+$ there are two coupled equations
whereas for $J^{\pi}=1^+$ their number is six but can actually be decoupled
into  2+4 \cite{MMB_These_01}.

$B_{ij}$  results from integration over $\varphi'$, the azimutal angle between
$\hat{k}$ and $\hat{k}'$, of more basic kernels $\kappa_{ij}$
\[ B_{ij}(k,u,k',u') = {1\over\pi^2} { {k'}^2 \over \epsilon_{k'}} \int_0^{2\pi}
\kappa_{ij}(k,u,k',u',\varphi')\;{F^2(Q^2)\over Q^2+\mu^2}\;{d\varphi'\over2\pi} \]
which have the general structure
\[ \kappa_{ij}(k,u,k',u',\varphi') = {1\over4} \sum_{\alpha\beta}
L_{\alpha\beta}\Black \; {\rm Tr} \left\{{\bf M}_{ij}^{\beta\alpha}\right\}\]
$F$ is a vertex form factor (FF) depending on $Q^2$,
$L_{\alpha\beta}$ is a tensor depending on the kind of coupling,
${\bf M}_{ij}^{\beta\alpha}$ is a contraction of a product of "$\gamma$-like" matrices
with a tensor  $\Pi_{\mu\nu}$ depending on the J$^{\pi}$ of the state.
\[ {\bf M}_{ij}^{\beta\alpha}= \sum_{\mu\nu}
 \;\Pi_{\mu\nu}\;\;
   S_{i}^{\mu}\; (\hat{k}_2+m)   \;     O_2^{\beta}   \; (\hat{k}^{'}_2+m) \;
{S'}_{j}^{\nu}\; (\hat{k}^{'}_1-m)  \;\bar{O}_1^{\alpha} \; (\hat{k}_1 -m)\]
where $\hat{k}=\gamma_{\mu}k^{\mu}$
and $ S_{i}^{\mu}$,${S'}_{i}^{\mu}$ are some spin operators
\cite{MCK_PRD_01,MMB_These_01}.

Kernels $\kappa_{ij}$ are in general calculated numerically.
Only for some special cases it is interesting to deal with analytical expressions,
which very soon turned out to be unreasonably lengthy.
As an example we give below the scalar kernel for J=0$^+$ states:
\begin{eqnarray*}\label{eqap1}
\kappa_{11}&=&-\alpha_s\pi \left\{
  (\varepsilon_{k}^2+\varepsilon_{k'}^2)(m^2-kk'\sin\theta\sin\theta'\cos\varphi')
+ 2\varepsilon_k \varepsilon_{k'}(\varepsilon_k\varepsilon_{k'}-kk'\cos\theta \cos\theta')\right\} \\
\kappa_{12}&=&-\alpha_s\pi m\left\{(\varepsilon_{k}^2-\varepsilon_{k'}^2)(k'\sin\theta' + k\sin\theta\cos\varphi')
 \right\}\\
\kappa_{21}&=&-\alpha_s\pi m\left\{(\varepsilon_{k'}^2-\varepsilon_{k}^2)(k\sin\theta +
k'\sin\theta'\cos\varphi')\right\} \\
\kappa_{22}&=& -\alpha_s\pi \left\{
(\varepsilon_{k}^2+\varepsilon_{k'}^2)(m^2\cos\varphi'-kk'\sin\theta\sin\theta')
+2\varepsilon_k\varepsilon_{k'} (\varepsilon_k\varepsilon_{k'}-k
k'\cos\theta \cos\theta'  ) \right\}
\end{eqnarray*} where we denote
$\alpha_{s}={g_s^2\over4\pi}$. Other analytical expressions have
also been obtained and can be found in \cite{MCK_PRD_BR_01,MCK_PRD_01,KMC_Prague_01,MMB_These_01}.

\subsection{Results}

In a series of works \cite{MC_PLB_00}-\cite{MMB_These_01}  still in progress we have separately
studied, coupling by coupling, $J^{\pi}=0^{\pm},1^{\pm}$ states in
the loosely bound (B$<<$m) and ultra relativistic ($B\sim m$)
limits. We will restrict ourselves in this contribution to scalar (S) and
pseudoscalar (PS) couplings with special emphasis in the stability problem -- are
equations soluble as they are provided by Quantum Field Theory or do they rather
need some regularization? -- and in their comparison with the non relativistic solutions.
\begin{figure}[htbp]
\begin{minipage}[htbp]{83mm}
\begin{center}
\mbox{\epsfxsize=8.cm\epsffile{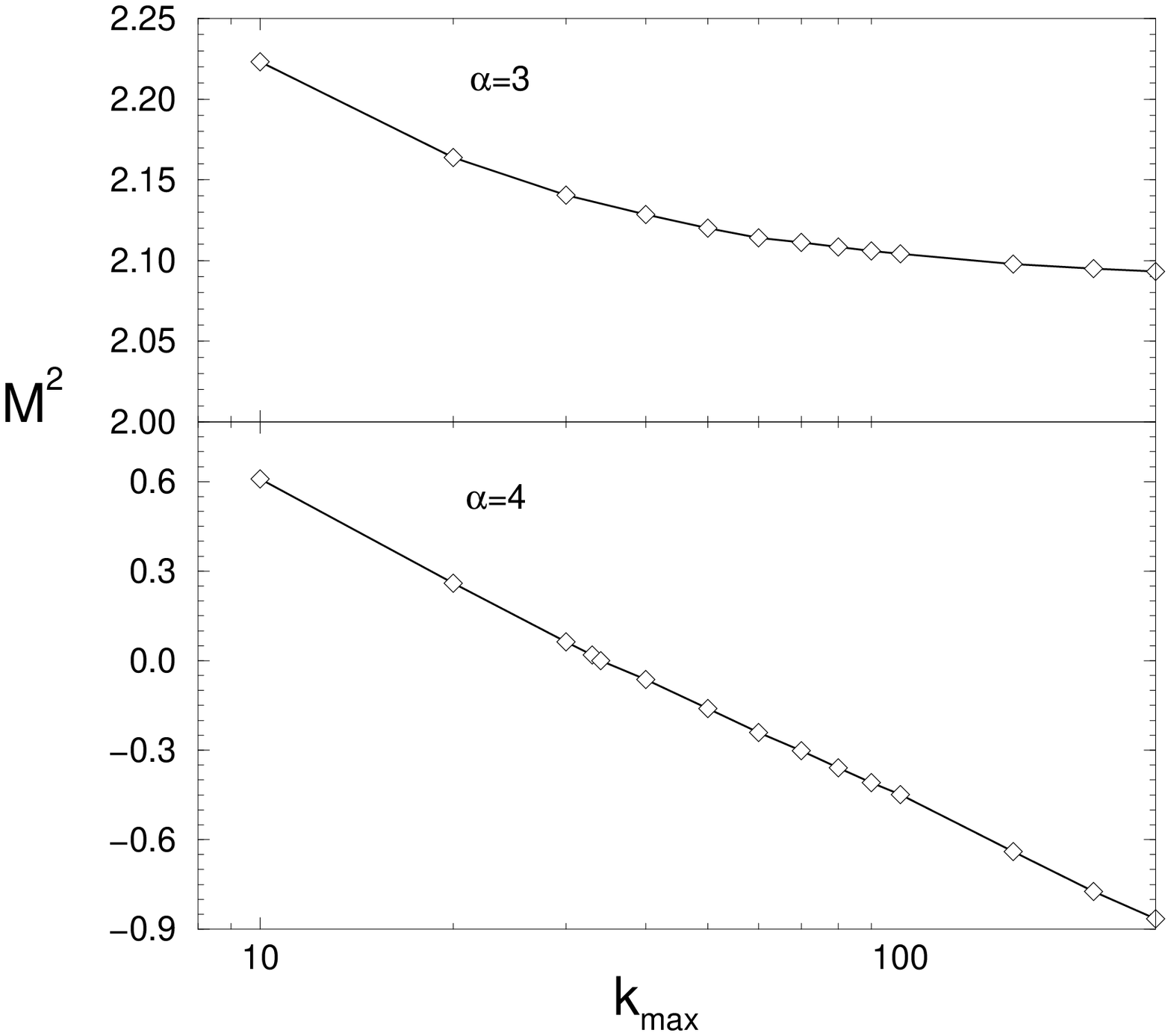}}
\caption{Cutoff dependence of $M^2$ in the $J=0^+$ state,
for two fixed coupling constant below and above the critical value.}\label{M2_kmax_mu=0.25_B22=0}
\end{center}
\end{minipage}
\hspace{0.2cm}
\begin{minipage}[htbp]{83mm}
\begin{center}
\mbox{\epsfxsize=8.cm\epsffile{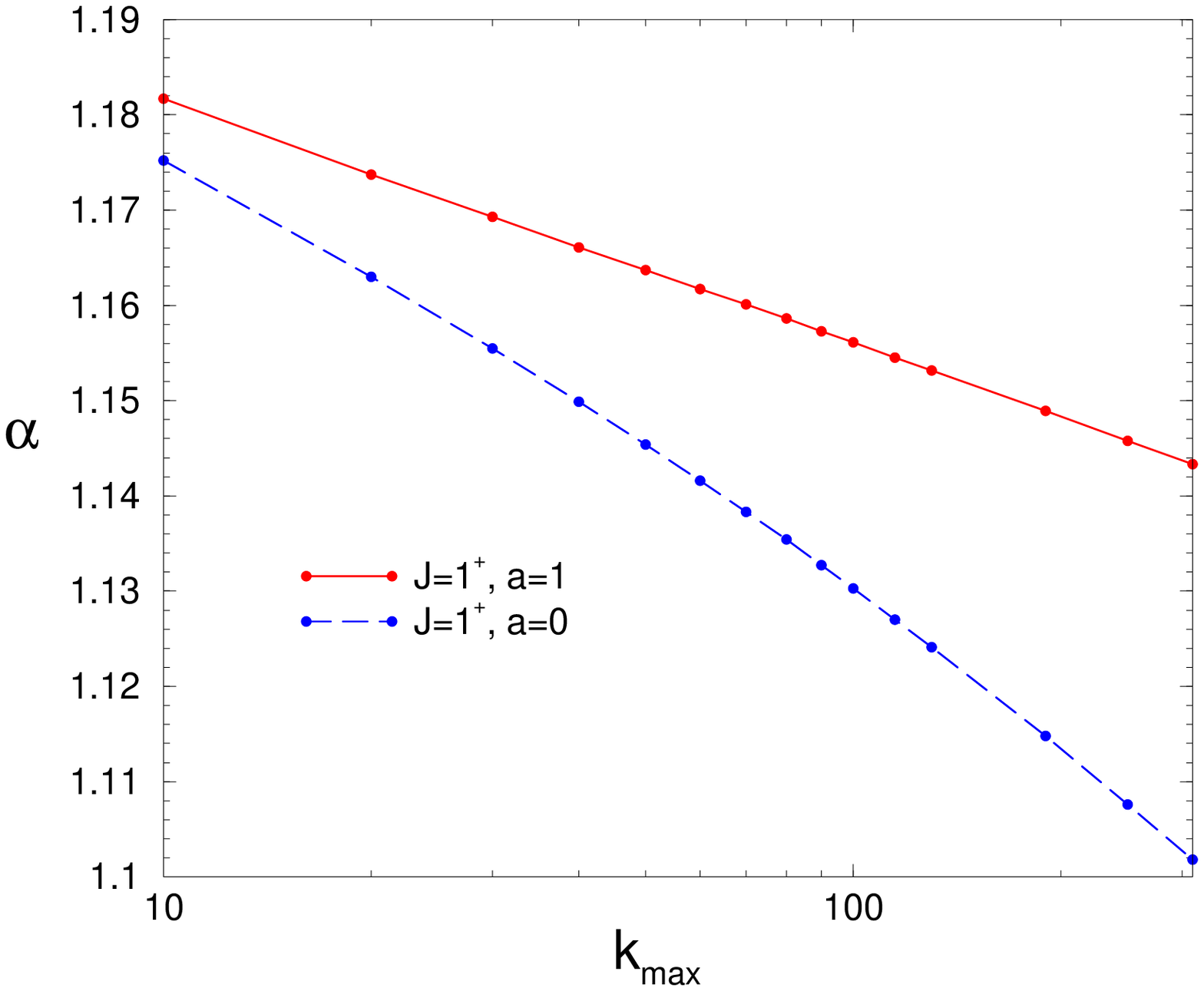}}
\caption{Cutoff dependence of the coupling constant, for $J=1^+$ states and
$B=0.05$.}\label{alpha_kmax_mu=0.25_J1a0a1}
\end{center}
\end{minipage}
\end{figure}

\subsubsection{The stability problem}

Let us solve the LFD equations in a compact domain $[0,k_{max}]$
\[ \left[M^2-4(k^2+m^2)\right] f(k,u) = \int_{0}^{k_{max}} dk' \int_{-1}^{+1 }du'\;
B(k,u,k',u') f(k',u') \]
and ask ourselves what happens with  $M^2$ when $k_{max}\rightarrow\infty$.
In case the solution exists, we will say that equations are stable.
In the opposite case, the theory would require some
regularization procedure, for instance by means of vertex form factors,
and it is then pertinent to inquire  how the solutions depend on them.
We have shown that the answer to this question
critically depends  on the type of coupling, on the $J^{\pi}$ quantum numbers of the state
and on the value of the coupling constant $\alpha$.

Consider first  the scalar coupling (Yukawa  model) for J=$0^+$.
We found \cite{MCK_PRD_01} the
existence of a critical coupling constant $\alpha_c\approx3.72$, below which
the solution exists, and above which the system "collapses".
This behavior is illustrated in figure  \ref{M2_kmax_mu=0.25_B22=0}
where we have plotted the $M$ dependence on $k_{max}$ for two
different values of $\alpha$.
We showed the asymptotical behavior of solutions to be
\[ f(k,u) = \frac{g(u)}{k^{2+\beta}} \]
with a relation $\beta(\alpha)$ provided by an eigenvalue equation
that allows a precise determination of the critical value $\beta(\alpha_c)=0$.
\begin{figure}[htbp]  
\begin{minipage}[htbp]{81mm}
\begin{center}\epsfxsize=80mm
\mbox{\epsffile{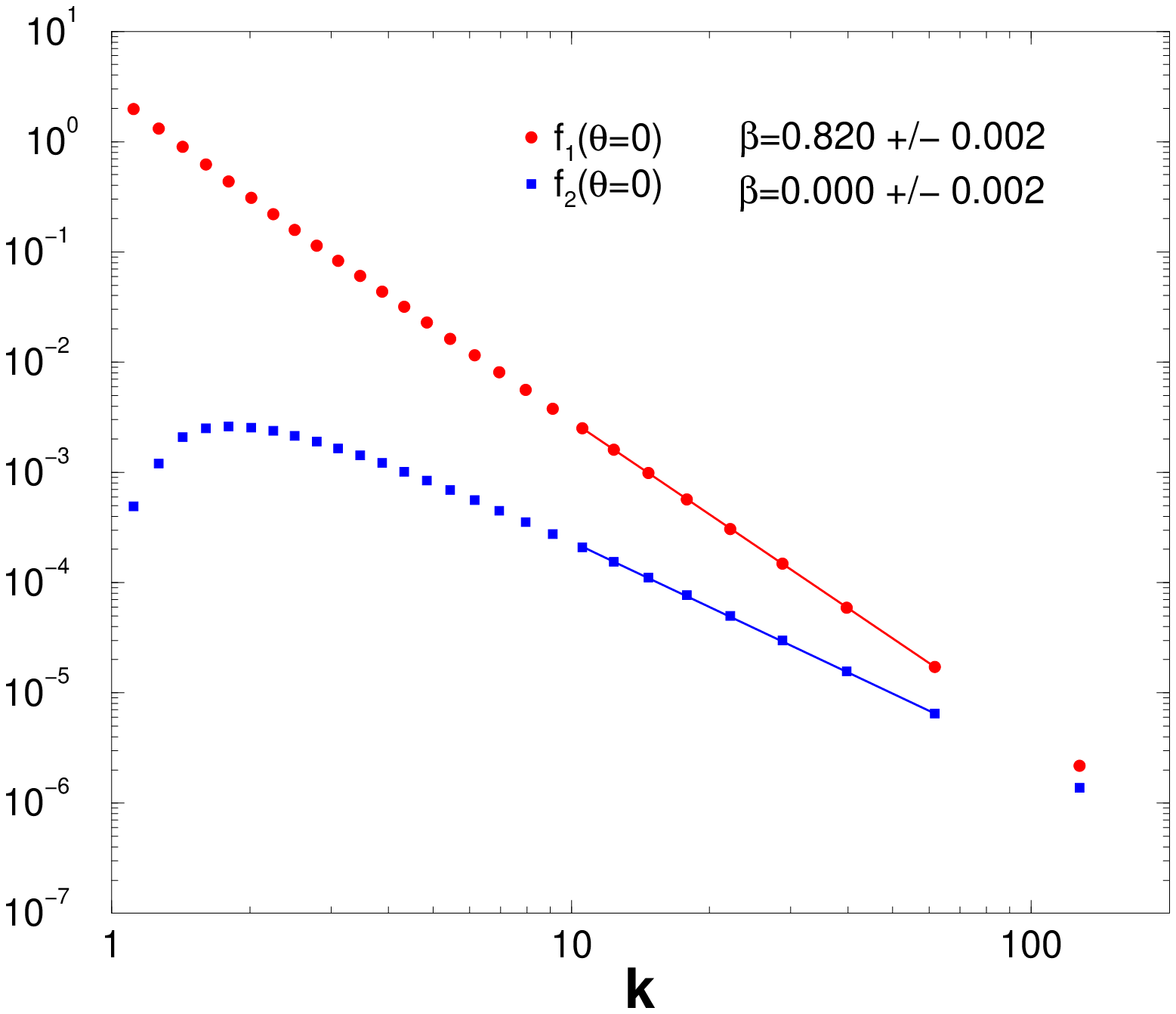}}\end{center}
\caption{Asymptotics of  $f_i$ for $J=0^+$ state with $B$=0.05, $\alpha$=1.096, $\mu$=0.25.
The power-law coefficients are $\beta_1=0.82$ and $\beta_2\approx0$}\label{wf_as_50}
\end{minipage}
\hspace{0.4cm}
\begin{minipage}[htbp]{81mm}   
\begin{center}\epsfxsize=80mm 
\mbox{\epsffile{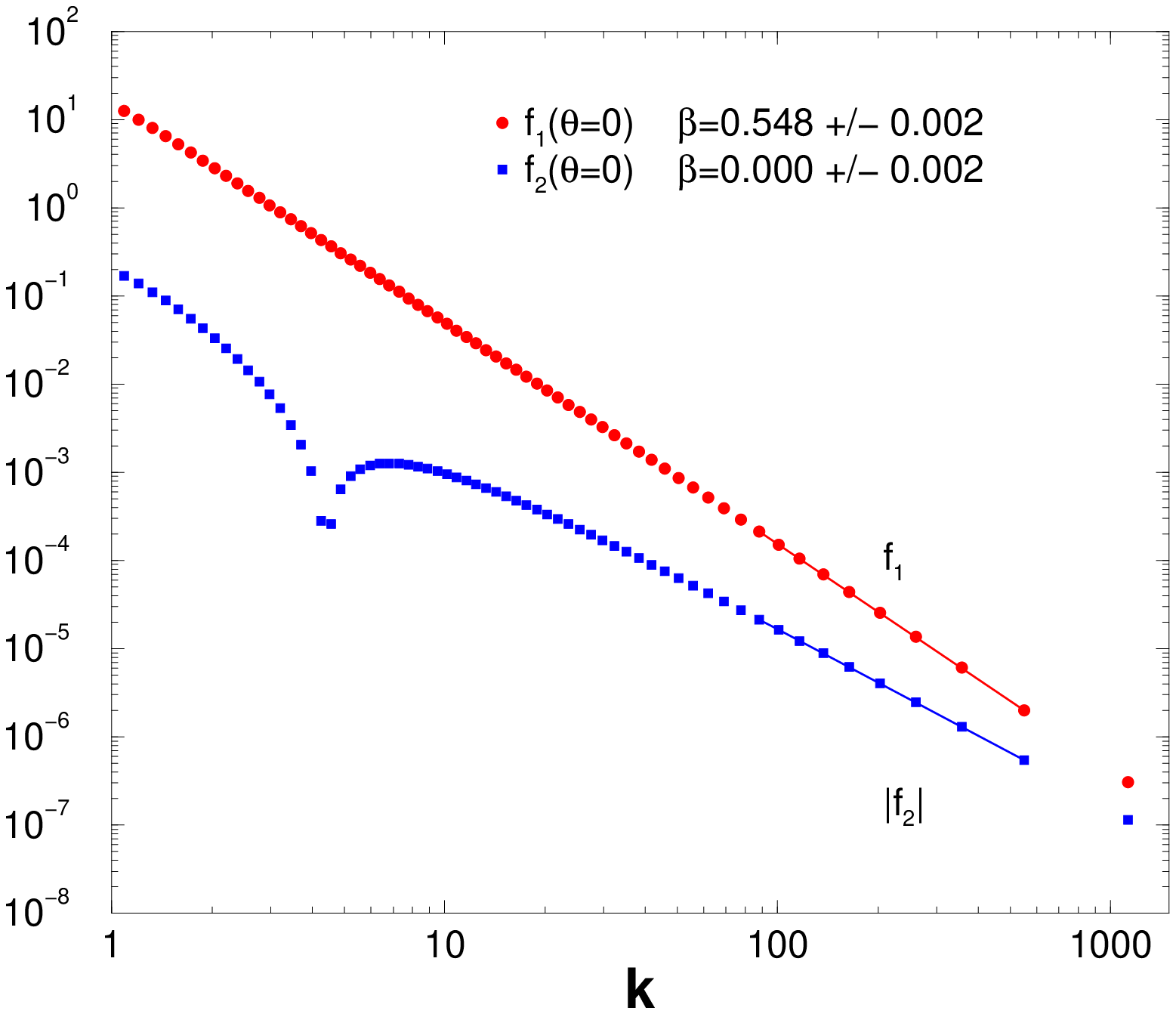}}\end{center}
\caption{Asymptotics of $f_i$ for $J=0^+$ state with $B=0.5$, $\alpha$=2.48, $\mu$=0.25.
The power-law coefficients are $\beta_1=0.55$ and $\beta_2\approx0$}\label{wf_as_500}
\end{minipage}
\end{figure}
This property was checked by a direct inspection of the numerical solutions of LFD equations
-- as it can be seen in figures \ref{wf_as_50} and \ref{wf_as_500} -- 
and turns to be very accurate.
It is worth noticing that -- at least in the framework of this model --
the coupling constant could be "measured" in the tail of the wavefunction!
The J=$1^+$ states, on the contrary,
do not present any stable solution without vertex form factors.
This can be seen in figure \ref{alpha_kmax_mu=0.25_J1a0a1}
where the divergency of the coupling constant as a function of $k_{max}$ is displayed
for the two angular momentum projections of a J=$1^+$ state with $B=0.05$.
One can also remark in this figure the non degeneracy
of the different projections due to the Fock space truncation; for the scalar coupling and moderate
values of $k_{max}\sim10$, it remains however less than one percent.

For pseudoscalar coupling, the stability analysis was performed using the same methods
than for the scalar one  \cite{MCK_LCM_01,MMB_These_01} and presents some peculiarities.
Quite surprisingly, the equations for J=$0^+$ states were found to be stable without any regularization.
The results were however strange in the sense that they lead to a quasidegeneracy of the
coupling constants for binding energies which can vary over all the physical range $[0,2m]$.
One gets for instance $\alpha=49.5$ for $B=0.001$
whereas $\alpha=48.6$ for a binding energy 500 times bigger $B=0.5$.
The origin of this anomaly was found to lie in the second channel equation ($\kappa_{22}$).
It has been understood analytically \cite{MCK_LCM_01} with a simple model but
leads to physically unacceptable results.

The case of other couplings (PV,V,T) has been examined in \cite{MMB_These_01}.
Corresponding equations results into more singular kernels which
all lead to unstable solutions without form factors.

\subsubsection{Comparison with non relativistic solutions}

We will present the results for $J=0^+$ states -- so with a
two-component wavefunction -- obtained with the scalar and
pseudoscalar coupling. In each of them, we will consider a loosely
bound (B=0.001) and a very relativistic (B=0.5) system. We have
fixed for all the calculations an exchanged mass $\mu=0.15$.
\begin{figure}[htbp]
\begin{minipage}[htbp]{80mm}
\begin{center}\mbox{\epsfxsize=7.9cm\epsffile{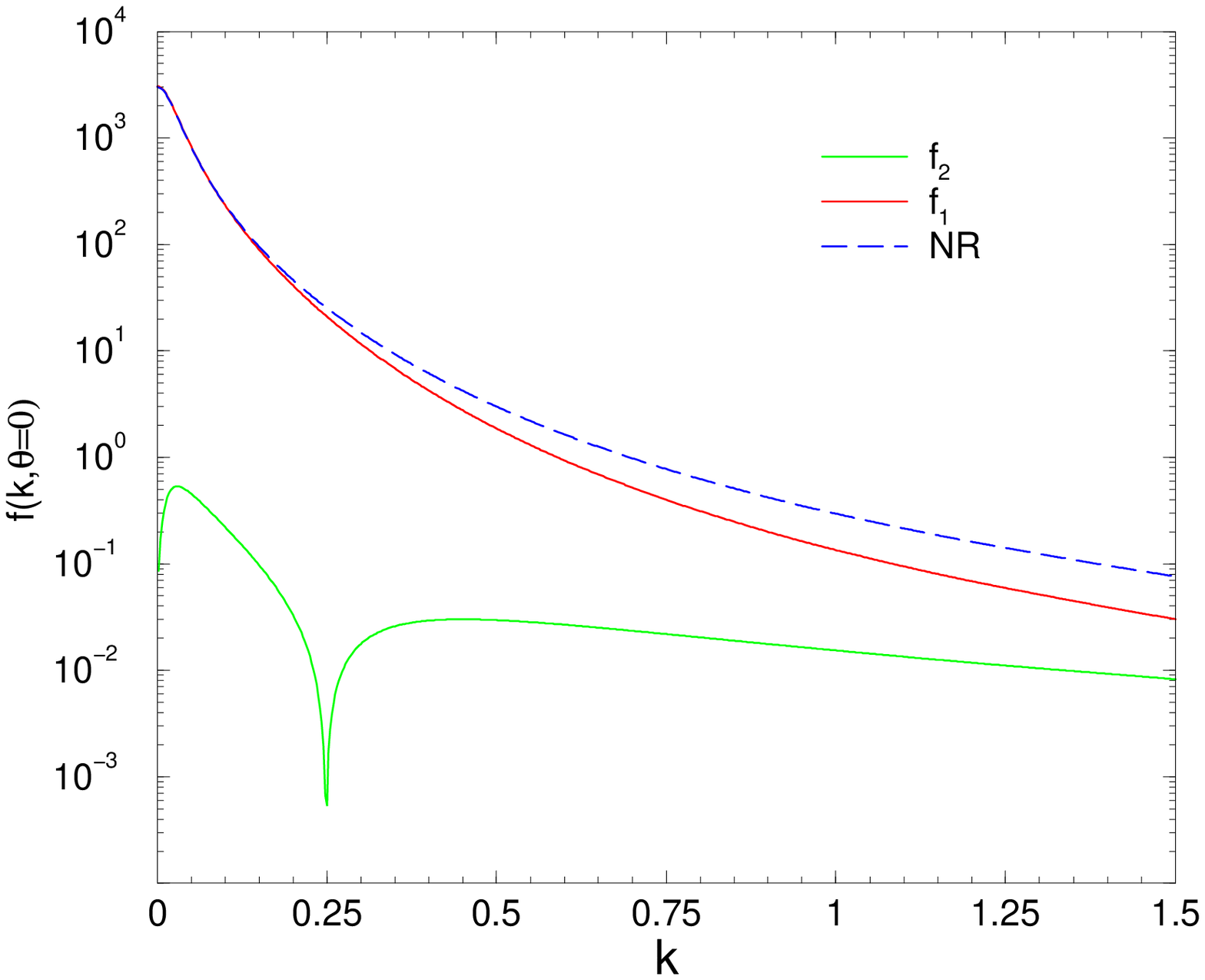}}\end{center}
\caption{LFD and non relativistic wavefunctions
 for J=$0^+$ state with $B=0.001$, $\mu=0.15$ in the Yukawa model}\label{S_f12_1_log_1MeV_sf}
\end{minipage}
\hspace{0.3cm}
\begin{minipage}[htbp]{80mm}
\begin{center}\mbox{\epsfxsize=7.9cm\epsffile{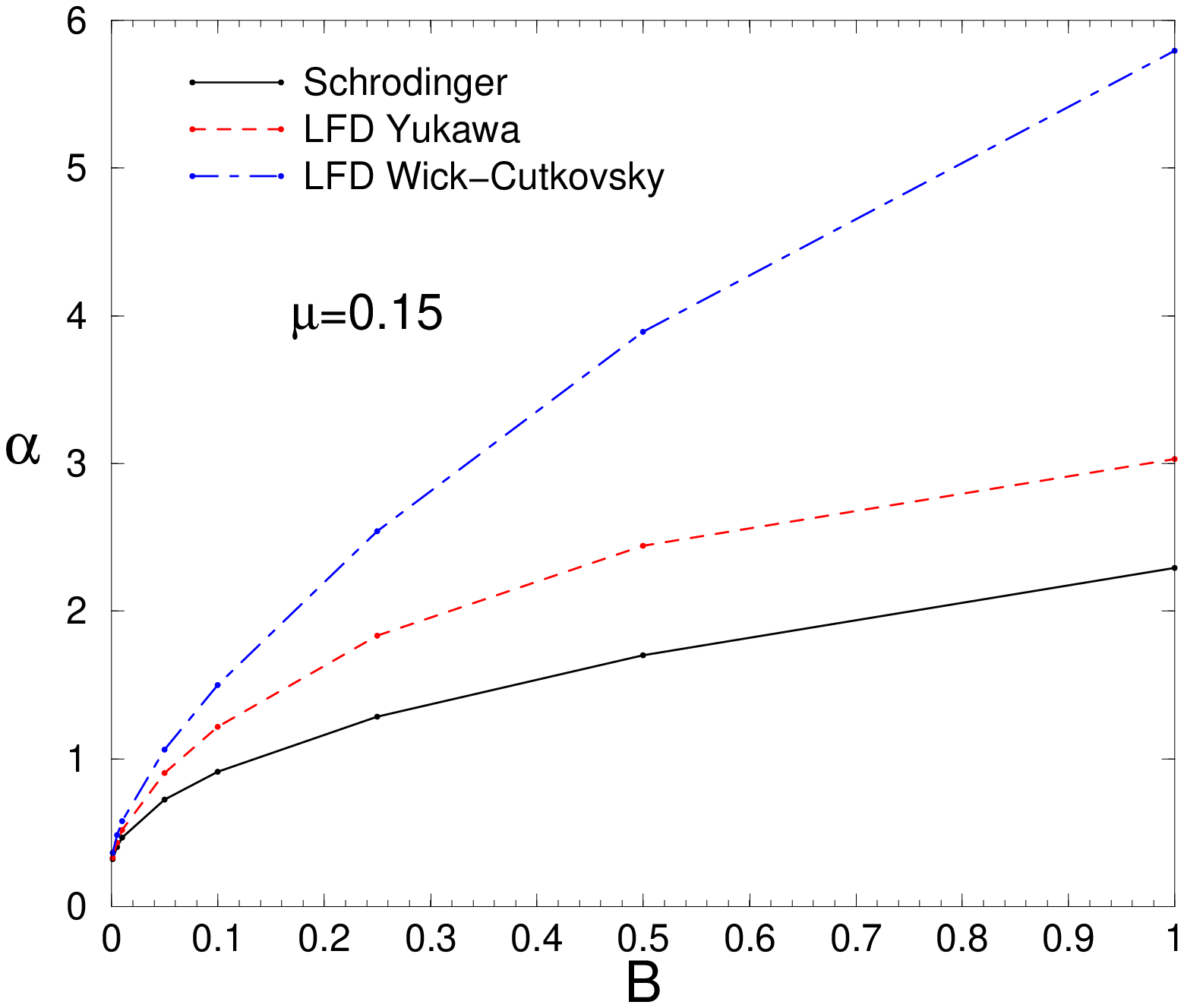}}\end{center}
\caption{Comparison between LFD (Yukawa model), WC and NR $B(\alpha)$ in $J=0^+$ state}\label{B_Y_WC_NR}\end{minipage}
\end{figure}

\paragraph{Scalar coupling}

$ $
\bigskip

For $B=0.001$, the LFD coupling constant is $\alpha_{LFD}$=0.331 whereas
the non relativistic  value is $\alpha_{NR}$=0.323.
Like in the Wick-Cutkosky model discussed in the last section
 -- scalar particles interacting by scalar exchange -- relativistic effects are repulsive.
They are responsible for only a 3\% difference in the coupling constants,  whereas in the purely scalar case
this difference is sensibly greater ($\alpha_{WC}$=0.364).
Wavefunctions -- suitably re-scaled --  are displayed in figure \ref{S_f12_1_log_1MeV_sf}.
One sees that component $f_1$ dominates over $f_2$ in all the interesting momentum range
and  that $f_2$ has a zero at $k\approx0.25$.
One can also remark that $f_1$ is very close to the NR wavefunction in
the small momentum region but it sensibly deviates with increasing $k$; for
$k\sim 1.5$ the difference represents more than one order of magnitude in the probability densities.
The coupling between the two LFD amplitudes has a very small (0.1\%) and attractive effect in the binding energy.

\begin{figure}[htbp]
\begin{minipage}[htbp]{80mm}
\begin{center}\mbox{\epsfxsize=7.9cm\epsffile{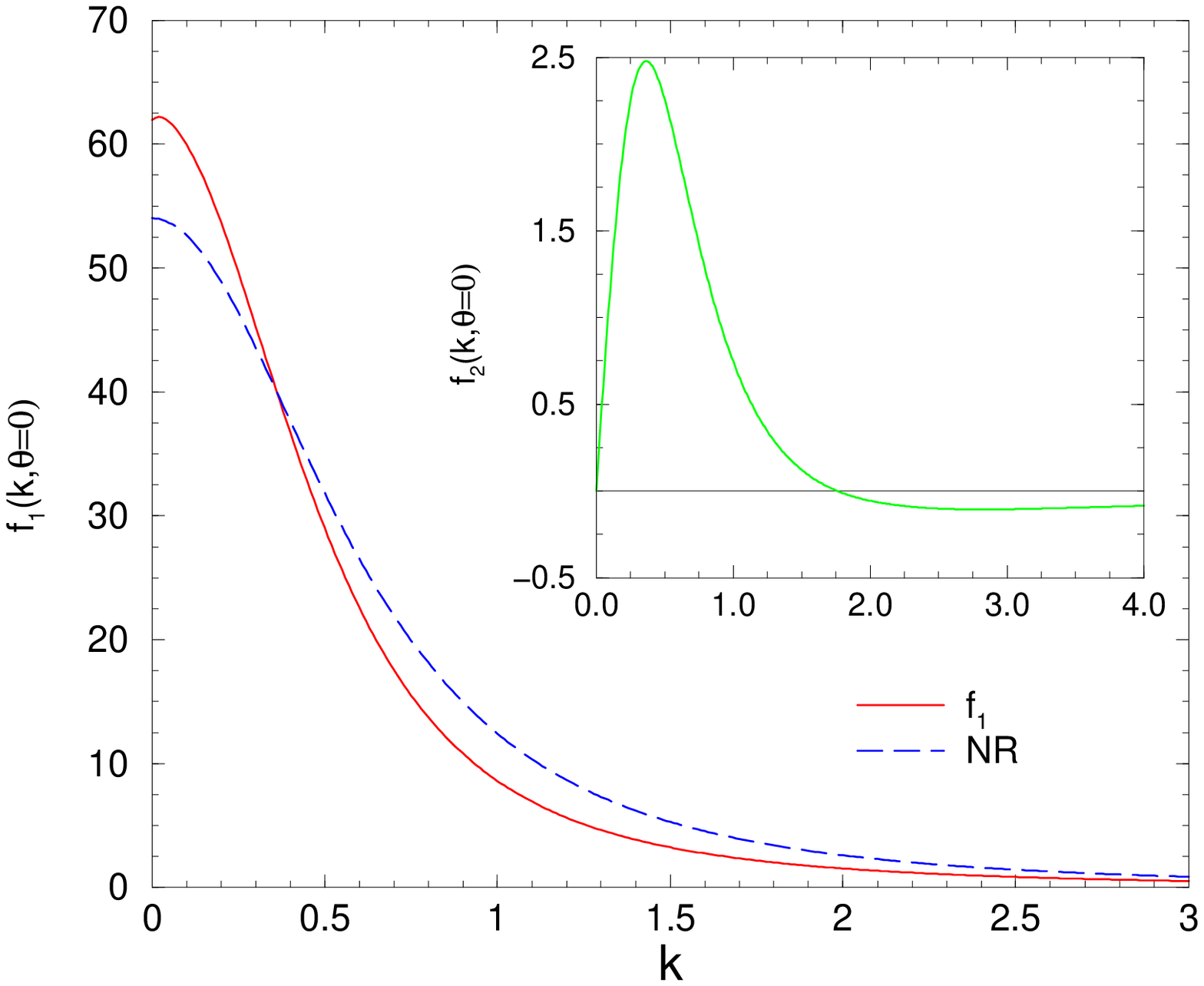}}\end{center}
\caption{LFD and non relativistic wavefunctions
 for J=$0^+$ state with $B=0.5$, $\mu=0.15$  in Yukawa model}\label{S_f12_2_500MeV_sf}
\end{minipage}
\hspace{0.2cm}
\begin{minipage}[htbp]{80mm}
\begin{center}\mbox{\epsfxsize=7.9cm\epsffile{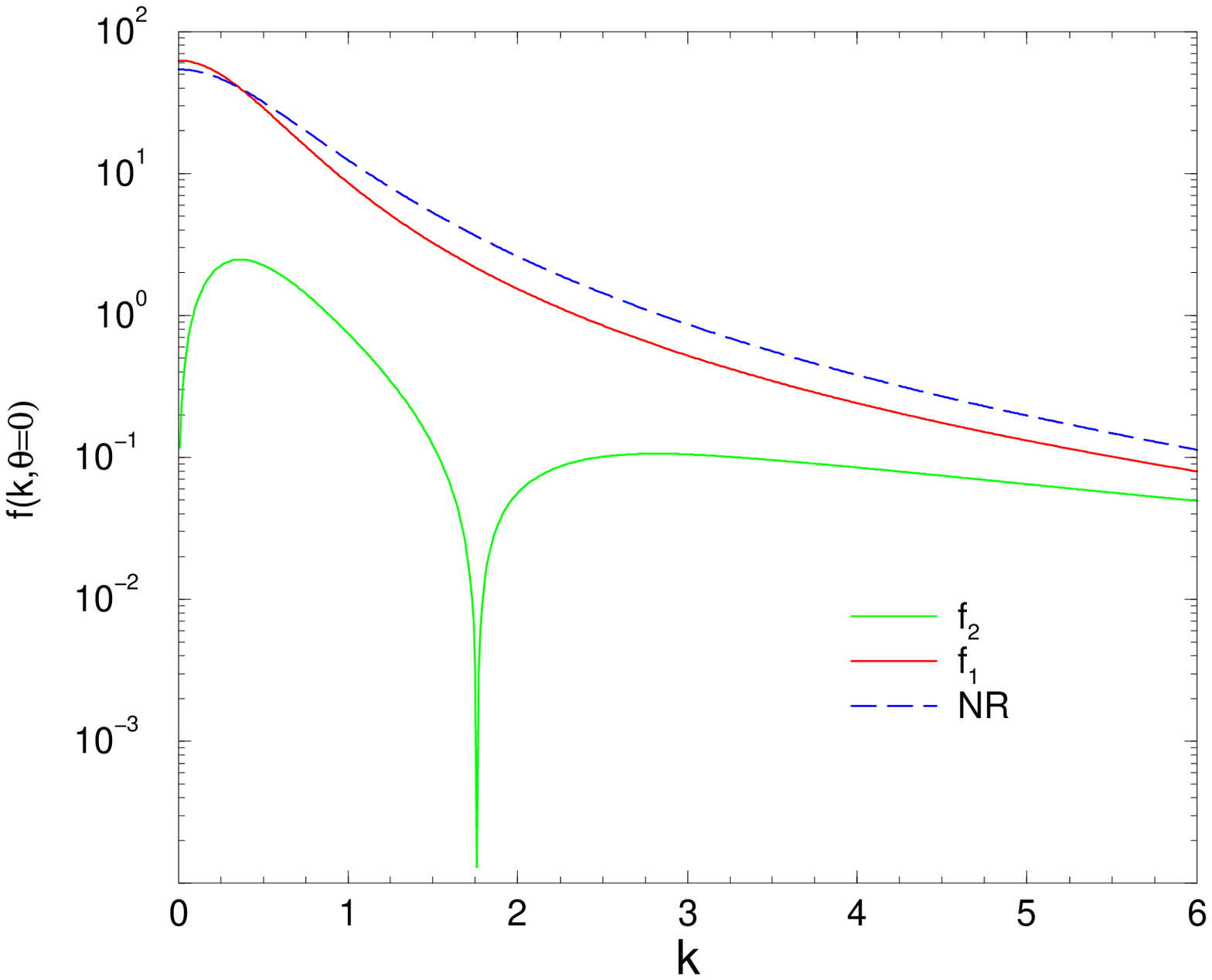}}\end{center}
\caption{LFD and non relativistic wavefunctions
 for J=$0^+$ state with $B=0.5$, $\mu=0.15$  in Yukawa model}\label{S_f12_2_log_500MeV_sf}
\end{minipage}
\end{figure}

In the strong binding limit (B=0.5) the situation is quite similar with enhanced relativistic effects.
One has $\alpha_{LFD}$=2.44 for $\alpha_{NR}$=1.71 and the differences in the wavefunctions -- displayed in figures
\ref{S_f12_2_500MeV_sf} and \ref{S_f12_2_log_500MeV_sf} -- are already visible at $k=0$ (figure
\ref{S_f12_2_500MeV_sf}).
One can see however in figure \ref{S_f12_2_log_500MeV_sf} that -- even
for deeply bound systems -- $f_1$ component still dominates over $f_2$.
To complete these results, we have displayed in figure \ref{B_Y_WC_NR} the LFD, NR and WC coupling
constants for different values of the binding energy.
One can see that the LFD results are systematically closer to the non relativistic values
than $\alpha_{WC}$ are, as if the fermionic character of the constituents
generates closer binding energies to the NR case, but larger differences in the high momentum
components of the wavefunction, due to the different asymptotics of interaction kernels.

\paragraph{Pseudoscalar  coupling}

$ $
\bigskip

For pseudoscalar coupling the situation is more involved.
First the use of vertex form factor -- though not necessary for
getting convergent solutions -- is essential to ensure physically meaningful results.
We have chosen, as in \cite{Bonn}, a form factor
\[ F(Q^2) =\left( { \Lambda^2-\mu^2\over \Lambda^2+ Q^2}\right)^n   \]
with $\Lambda=1.3$ and $n=1$.

\begin{figure}[htbp]
\begin{minipage}[htbp]{80mm}
\begin{center}
\mbox{\epsfxsize=7.9cm\epsffile{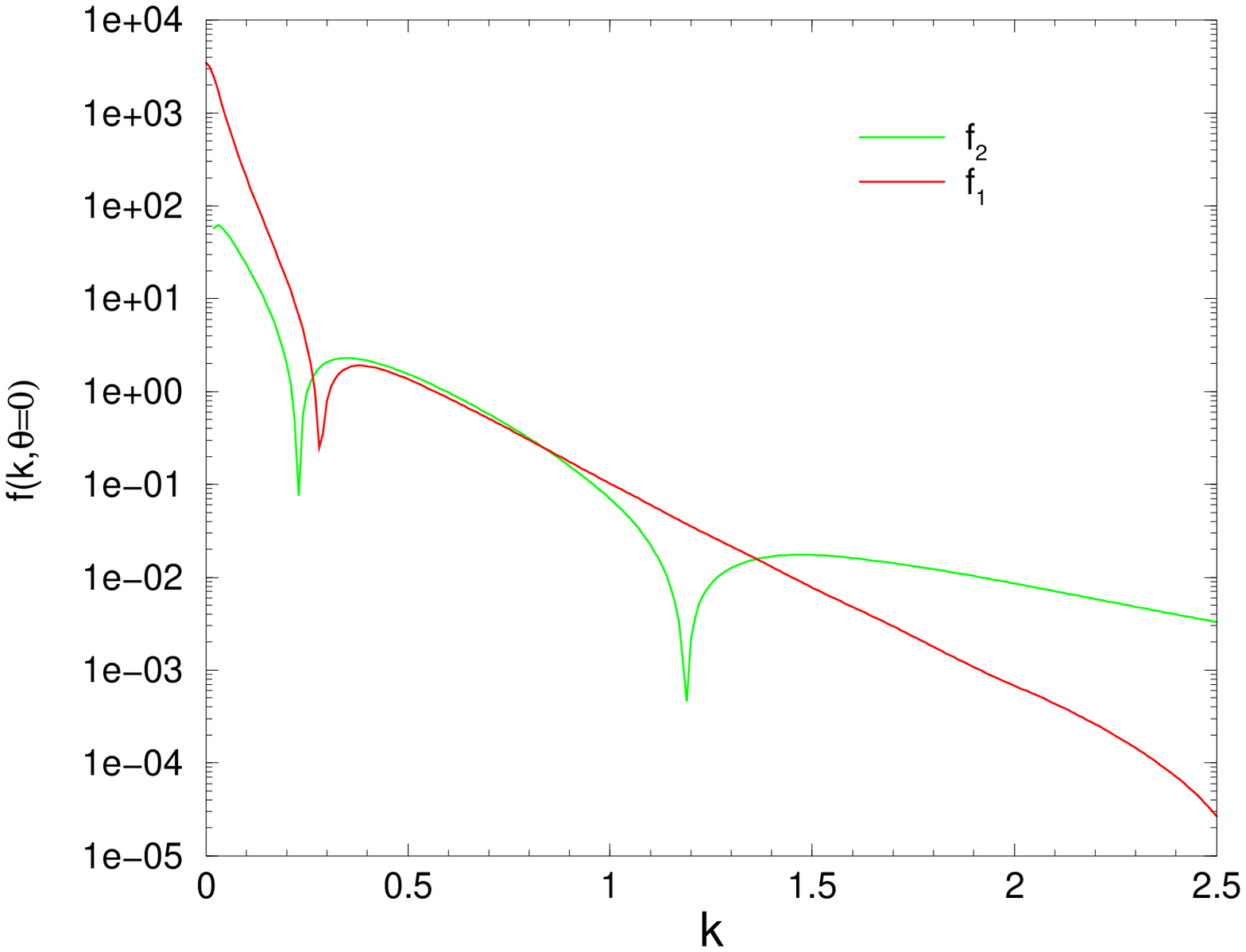}}
\caption{LFD and non relativistic wavefunctions
 for J=$0^+$ state with $B=0.001$, $\mu=0.15$ and PS coupling}\label{PS_f12_1_I=1_1MeV_ff_log}
\end{center}
\end{minipage}
\hspace{0.2cm}
\begin{minipage}[htbp]{80mm}
\begin{center}
\mbox{\epsfxsize=7.9cm\epsffile{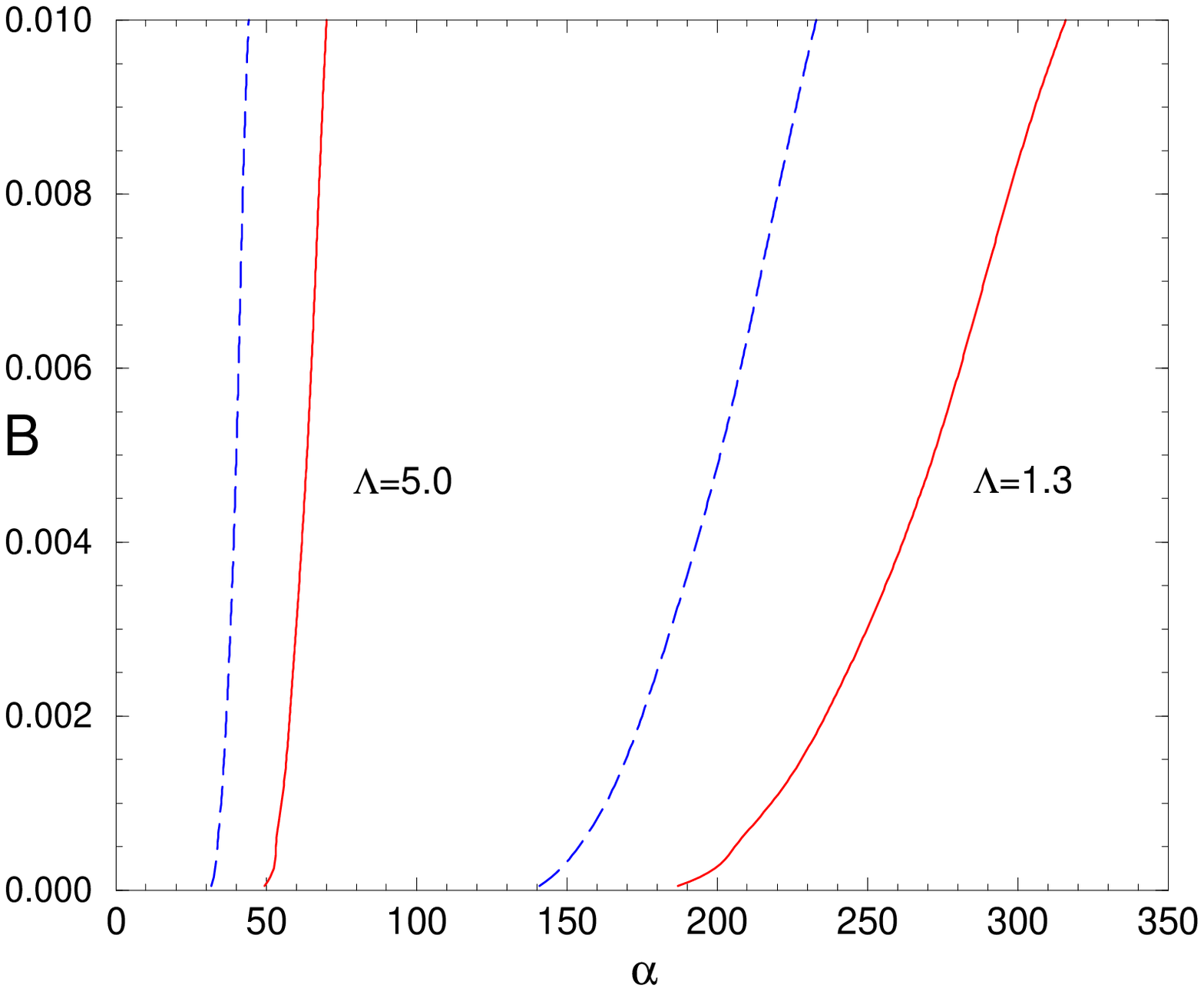}}
\caption{Zero binding limit in LFD (solid) and Schrodinger (dashed) equations
with PS coupling and different form factors ($\mu=0.5$)}\label{B=0_ps}
\end{center}
\end{minipage}
\end{figure}

In the weak binding limit (B=0.001) one has  $\alpha_{LFD}=190$
and $\alpha_{NR}=166$, a repulsif effect much stronger (15\%)
than in the scalar coupling. Corresponding wavefunctions are shown
in figure \ref{PS_f12_1_I=1_1MeV_ff_log}. One can see that
$f_2\approx f_1$ at $k\sim0.3$ and dominates above $k$=1.
It is worth noticing the dramatic influence of the form factor in all
these calculations: one has for instance $\alpha_{LFD}$=103  for
$\Lambda=5$ and $\alpha_{LFD}$=1725 for $\Lambda$=0.3 !

The coupling between the two components $f_i$ is also very important.
By switching off the non diagonal
kernels $B_{12}=B_{21}=0$ the coupling constant moves
from $\alpha_{LFD}=190$ to $\alpha_{LFD}=251$. It has thus
an attractive effect which tends to minimize the difference between LFD and NR results.

Quite surprisingly, in the strong binding limit (B=0.5) we have
found $\alpha_{LFD}$=1462 and $\alpha_{NR}$=3065. Relativistic
effects become now strongly attractive
($\alpha_{LFD}<\alpha_{NR}$). An essential part of  this
attraction is due to coupling $f_1-f_2$ of the two components in
the LFD wavefunction. By performing one channel calculations one
has indeed $\alpha_{LFD}$=3001, what represents a strong reduction
in the effect though it remains slightly attractive. We have
checked if this happens for different values of the exchange mass
$\mu$. For the same binging energy $B=0.5$ and $\mu=0.5$ one has
$\alpha_{LFD}$=1728 and $\alpha_{NR}$=1400, repulsive once again.
This told us the difficulty of talking about relativistic effects
in general. They turn to depend not only on the kind of coupling
but also on the binding energy of the system and, furthermore, on
the mass of the exchanged particle.

The preceding results show a qualitative difference between
S and PS cases. Pseudoscalar coupling is by construction relativistic:
small and large spinor components are mixed to the first order.
Moreover the coupling between $f_1$ and $f_2$ is essential even for very weakly bound systems.
It is so interesting to study in this case the zero energy limit of the LFD and
compare with the non relativistic results. We understand by that the static potential,
with the same form factor, including delta function term.
The results, displayed in figure \ref{B=0_ps}, correspond to $\mu=0.5$
and two different cutoff parameters $\Lambda$
in the form factors. They show the same behavior that was found in the scalar case,
i.e. that relativistic and non relativistic approaches did not coincide even for
systems with zero binding energies as fas as they interact with massive exchanges.

\section{Summary}

We have presented the main ideas of the Light-Front Dynamics formalism in its explicitly
covariant version.
When applied to nucleon-nucleon system with perturbative wavefunctions,
they led to results which are in
good agreement with the TJNAF data,
even for momentum transfers larger than the constituent masses.
Recently, an interesting application of the Light-Front nucleon-nucleon wavefunctions
was done to successfully describe the two-body correlations in light nuclei \cite{Antonov_02}.
These facts indicate the ability of this approach
in describing composite relativistic systems and claim for further developments on this theory.

This work was pursued by calculating
the Light-Front solutions in the ladder approximation for two-scalar and two-fermion systems
and the main results obtained up to now have been summarized in this contribution.
Among them we would like to stress the points that follow.

In the scalar case, we have found that the inclusion
of relativity has a dramatic repulsive effect on binding energies
even for systems with very small ${<k>\over m}$ values.
The effect is specially relevant when using a scalar model for
deuteron: its binding energy is shifted from 2.23 MeV down to 0.96 MeV.
This can be corrected by a 10\% decrease of the repulsive coupling constant, what indicates
the difficulty to determine beyond this accuracy the value of strong coupling constants
within a non relativistic framework.

Light-Front wavefunctions strongly differ from their non
relativistic counterparts if they are calculated using the same
coupling constant. Once the interaction parameters are readjusted
to get the same binding energy, both solutions become closer in the
small momentum region but their deviations are sizeable at $k\sim m$.

The relativistic effets are shown to be induced mainly by the
additional terms appearing in the interaction kernel.
Kinematical corrections have only a small influence on the binding energy.

The Light-Front results are found to be quite
close to those provided by Bethe-Salpeter equation, for a wide range
of coupling constant, despite the different physical input in their ladder kernel.

Even in the zero binding limit, Light-Front and Schrodinger
solutions differ ($\mu\ne0$) both for scalars and fermions.
This leads to the conclusion that
such systems cannot be properly described by using a non relativistic dynamics.

For the Yukawa model, we found a critical coupling constant
below which the J=0$^+$ solutions are stable without any regularization.
For the pseudoscalar coupling, these solutions are
also stable for all values of the coupling constant but
lead to a quasidegenerate $\alpha(B)$ relation.
The J=$1^+$ states are on the contrary unstable for both couplings.

Pseudoscalar coupling shows large deviations with respect to
 non relativistic solution.
Wavefunction is dominated by relativistic components at $k<<m$,
even for weakly bound systems. However form factors play there a
determinant role.

The question about relativistic effects has no simple answer.
The consequences of implementing the Lorentz invariance in
a quantum mechanical description of a system are different, following:
the nature of its constituents, the kind of interaction,
the quantum numbers of the state, its binding energy,
and even the mass of the exchanged particle!
There are no simple recipes to perform {\it a priori} evaluations.

\bigskip
{\bf Acknowledgements:} This work is partially supported by the
Russian-French  research contract PICS No. 1172. Numerical
calculations were performed  at CGCV (CEA) and
IDRIS (CNRS). We thank the staff members  of these two organizations for their constant support.

\footnotesize{

}

\end{document}

%% file: CrayolaColors.tex
\def\Black{} 

%% file: Varna_paper.bbl
\begin{thebibliography}{10}
\bibitem{GVO_Revue_01} M. Gar\c{c}on, J.W. Van Orden;  Adv. Nucl. Phys. 26 (2001) 293; nucl-th/0102049
\bibitem{Gross} F. Gross, R. Gilman, nucl-th/0110015; R. Gilman, F. Gross, nucl-th/0111015
\bibitem{CK_NPA581_95}    J. Carbonell, V.A. Karmanov, Nucl. Phys. {\bf A581} (1995) 625.
\bibitem{CK_NPA589_95}    J. Carbonell, V.A. Karmanov, Nucl. Phys. {\bf A589} (1995) 713.
\bibitem{CDMK_98}         J. Carbonell, B. Desplanques, J.F. Mathiot, V.A. Karmanov, Phys. Rep. {\bf 300} (1998) 218
\bibitem{CK_EPJA_99}      J. Carbonell and V.A. Karmanov, Eur. Phys. J. {\bf A6} (1999) 9.
\bibitem{Bonn} R. Machleidt, K. Holinde, C. Elster,  Phys. Rep. {\bf 149}  (1987) 1
\bibitem{Antonov_02} A.N. Antonov et al, Phys. Rev, {\bf C} (2002); nucl-th/0106044
\bibitem{MC_PLB_00}    M. Mangin-Brinet, J. Carbonell, Phys. Lett. {\bf B474}, (2000) 237
\bibitem{KCM_Taiw_01}  V.A. Karmanov, J. Carbonell, M. Mangin-Brinet, Nucl. Phys. {\bf A684} (2001) 366c
\bibitem{MC_Evora_01}  M. Mangin-Brinet, J. Carbonell, Nucl. Phys. {\bf A689} (2001) 463c
\bibitem{MCK_Heid_00}  M. Mangin-Brinet, J. Carbonell, V.A. Karmanov, Nucl. Phys. {\bf B90} (2000) 123
\bibitem{MCK_PRD_BR_01}  M. Mangin-Brinet, J. Carbonell, V.A. Karmanov, Phys. Rev. {\bf D64}, (2001) 125005;hep-th/0102068
\bibitem{MCK_PRD_01}     M. Mangin-Brinet, J. Carbonell, V.A. Karmanov, Phys. Rev. {\bf D64}, (2001) 027701;hep-th/0107235
\bibitem{KMC_Prague_01}  V.A. Karmanov, J. Carbonell, M. Mangin-Brinet; to appear in ; hep-th/0107237
\bibitem{MCK_LCM_01}     M. Mangin-Brinet, J. Carbonell, V.A. Karmanov, to appear in Nucl. Phys. (2002);  hep-th/0112017 
\bibitem{KCM_LCM_01}     V.A. Karmanov, J. Carbonell, M. Mangin-Brinet, to appear in Nucl. Phys. (2002); nucl-th/0112005 
\bibitem{MMB_These_01} M. Mangin-Brinet, Th\`ese Universit\'e de Paris VII (2001)
\bibitem{VAK} V.A. Karmanov, JETP {\bf 56} (1982) 1; JETP {\bf 44} (1976) 210; Sov. J. Part. Nucl. {\bf 19}  (1988) 228
\bibitem{VAK_Mitra} V.A. Karmanov, Quantum Field Theory. A Twentieth Century Profile. 
         Ed. A.N. Mitra, Hindustanian Book Agency (India), (2000) 795
\bibitem{Dirac_RMP_49}        P.A.M. Dirac, Rev. Mod. Phys. {\bf 21} (1949) 392
\bibitem{kadysh} V.G. Kadyshevsky, ZhETF {\bf 46} (1964) 645, 872 [JETP {\bf 19} (1964) 443, 597]; Nucl. Phys. {\bf B6} (1968) 125
\bibitem{WEINBERG_66} S. Weinberg, Phys. Rev. {\bf 150} 1313 (1966)
\bibitem{CM_69}       S. Chang, S. Ma,     Phys. Rev. {\bf 180} 1506 (1969)
\bibitem{Coester}    F. Coester, Prog. Part. Nucl. Phys. {\bf 29} 1 (1992)
\bibitem{Glazek}     S. Glazek et al, Phys. Rev. {\bf D47} (1993) 1599
\bibitem{Ji}         Ch.-R. Ji, Phys. Lett. B322 (1994) 389  
\bibitem{Fuda}       M.G. Fuda, Y. Zhang, Phys. Rev. {\bf C51} (1995) 23
\bibitem{Burkart}    M. Burkardt, Adv. in Nucl. Phys. Vol. {\bf 23}  (1996) 1
\bibitem{Brodsky}    S.J. Brodsky, H.-C. Pauli and S.S. Pinsky,  Phys. Rep., {\bf 301} (1998) 299
\bibitem{Bakker}     N. Schoonderwoerd, B.L.G. Bakker, V.A. Karmanov, Phys. Rev. {\bf C58} (1998) 3093
\bibitem{Frederico}  J.H.O. Sales, T. Frederico, B.V. Carlson, P.U. Sauer, Phys. Rev. {\bf C61} (2000) 04003
\bibitem{Hiller}     J.R. Hiller, Nucl. Phys. {\bf B90} (2000) 170 
\bibitem{Miller}     J.R. Cooke, G.A. Miller; nucl-th/0112037
\bibitem{KS_94}   V.A. Karmanov, A.V. Smirnov, Nucl. Phys {\bf A575} (1994) 520
\bibitem{WC_54} G.C. Wick, Phys. Rev. {\bf 96} (1954) 1124; R.E. Cutkosky Phys. Rev. {\bf 96} (1954) 1135
\bibitem{CMP_PRC61_00} J.R. Cooke, G.A. Miller, D. Phillips, Phys. Rev. {\bf C61} (2000) 064005
\bibitem{FFT_73} G. Feldman, T. Fulton, J. Townsend, Phys. Rev. {\bf D7} 1814 (1973)
\bibitem{MT_69} R.A. Malfliet, J.A. Tjon, Nucl. Phys. {\bf A127} (1969) 161
\bibitem{Desplanques} A. Amghar, B. Desplanques, L. Theussl, Nucl. Phys. {\bf A694} (2001) 439
\end{thebibliography}
